\documentclass[12pt]{article}

\usepackage{graphicx,amssymb,url,mathrsfs, amsmath}
\usepackage{eucal}
\usepackage{wrapfig}
\usepackage{boxedminipage}
\usepackage{setspace}
\usepackage{subfigure}
\usepackage{amsxtra,amstext,latexsym,dsfont}

\def\a  {\alpha}       \def\b  {\beta}         \def\g  {\gamma}
\def\G  {\Gamma}       \def\d  {\delta}        \def\D  {\Delta}
\def\e  {\epsilon}        \def\k  {\kappa}
\def\l  {\lambda}             \def\m  {\mu}
          \def\s  {\sigma}        
\def\t  {\tau}                 
   \def\w  {\omega}

 \newcommand{\cald}{\mbox{${\cal D}$}}

 \newcommand{\caln}{\mbox{${\cal N}$}}
\newcommand{\calo}{\mbox{${\cal O}$}}

 \newcommand{\calv}{\mbox{${\cal V}$}}

\def\IR{{\hbox{{\rm I}\kern-.2em\hbox{\rm R}}}}
\def\IB{{\hbox{{\rm I}\kern-.2em\hbox{\rm B}}}}
\def\IN{{\hbox{{\rm I}\kern-.2em\hbox{\rm N}}}}
\def\IC{\,\,{\hbox{{\rm I}\kern-.59em\hbox{\bf C}}}}
\def\IZ{{\hbox{{\rm Z}\kern-.4em\hbox{\rm Z}}}}
\def\IP{{\hbox{{\rm I}\kern-.2em\hbox{\rm P}}}}
\def\IH{{\hbox{{\rm I}\kern-.4em\hbox{\rm H}}}}
\def\ID{{\hbox{{\rm I}\kern-.2em\hbox{\rm D}}}}

\def\be{\begin{equation}}
\def\ee{\end{equation}}
\def\ba{\begin{eqnarray}}
\def\ea{\end{eqnarray}}

\def\half{\frac{1}{2}}

\newcommand{\inv}[1]{\frac{1}{#1}}

\def\ra{\rightarrow}  

\newcommand{\dint}[2]{\int_{#1}^{#2}\!\!}

\def\dell{\partial}

\newcommand{\bra}[1]{\mbox{$\langle #1 |$}}
\newcommand{\ket}[1]{\mbox{$| #1 \rangle$}}

\def\Tr{{\rm tr}\,}

\def\nn{\nonumber}
\def\ea{{\it et al}. }

\def\De{\textrm{D8}}
\def\DeB{\overline{\textrm{D8}}}

\newcommand{\Ukk}{U_{\rm KK}}
\newcommand{\Mkk}{M_{\rm KK}}
\newcommand{\wt}{\widetilde}

\newcommand{\nw}{{{ \mathfrak w}}}
\newcommand{\nq}{{{ \mathfrak q}}}
\newcommand{\nd}{{{ \mathbf{d}  }}}

\begin{document}

\begin{titlepage}


\begin{center}
{\large \bf Baryonic Response of Dense Holographic QCD} \\
\vspace{10mm}
  Keun-Young Kim$^a$ and Ismail Zahed$^a$\\
\vspace{5mm}
$^a$ {\it Department of Physics and Astronomy, SUNY Stony-Brook, NY 11794}\\
\vspace{10mm}
\end{center}
\begin{abstract}
The response function of a homogeneous and dense hadronic system to a
time-dependent (baryon) vector potential is
discussed for holographic dense QCD (D4/D8 embedding) both in the confined
and deconfined phases. Confined holographic QCD is an uncompressible
and static baryonic insulator at large $N_c$ and large $\l$, with a gapped vector
spectrum and a massless pion. Deconfined holographic QCD is a diffusive
conductor with restored chiral symmetry and a gapped transverse baryonic current.
Similarly, dense D3/D7 is diffusive for any non-zero temperature at large $N_c$
and large $\l$. At zero temperature dense D3/D7 exhibits a baryonic longitudinal
visco-elastic mode with a first sound speed $1/\sqrt{3}$ and a small width due
to a shear viscosity to baryon ratio $\eta/n_B=\hbar / 4$. This mode is turned
diffusive by arbitrarily small temperatures, a hallmark of holography.
\end{abstract}

\end{titlepage}

\renewcommand{\thefootnote}{\arabic{footnote}}
\setcounter{footnote}{0}

\tableofcontents


\section{Introduction}

Solids respond to external stress elastically through their bulk
and shear modulii $K$ and $\mu$ respectively, with almost zero
dissipation. Liquids on the other hand, follow the lore of
hydrodynamics with bulk and shear viscosities $\xi$ and $\eta$
accounting for dissipation. In contrast to the solid, the shear
modulus vanishes in the liquid. The bulk modulus does not.

This remarkable difference between solid and liquid disappears
when the stress is time-dependent. Indeed, for a stress of finite
frequency $\omega$ a liquid has a non-zero shear modulus much like
the solid. In the long-wavelength limit, the dual description of
a solid or a liquid follows from the visco-elastic equations with
complex and frequency dependent elastic constants as we detail
below. In this paper we will explore some of these ideas in the
context of the AdS/CFT correspondence by analyzing the baryonic
response functions at finite density for both D4/D8 and D3/D7
embeddings.

The AdS/CFT approach~\cite{Maldacena} provides a useful
framework for discussing large $N_c$ gauge theories at strong
coupling $\lambda=g^2N_c$. The model suggested by Sakai and
Sugimoto (SS) \cite{Sakai1} offers a specific holographic
realization of hQCD that includes $N_f$ flavors and is chiral. For
$N_f\ll N_c$, chiral QCD is obtained as a gravity dual to $N_f$
$\De$-$\DeB$ branes embedded into a $D4$ background in 10
dimensions where supersymmetry is broken by the Kaluza-Klein (KK)
mechanism. The SS model yields a holographic description of
hadrons in the vacuum~\cite{Sakai1, Sakai2, Baryons1, Baryons2, KZ, Baryons3},
at finite temperature~\cite{FiniteT} and finite baryon
density~\cite{Finitemu, KSZ2,KSZ3, KSZ4}.

Hot and dense hadronic matter in QCD is difficult to track from
first principles in current lattice simulations owing to the sign
problem. In large $N_c$ QCD baryons are solitons and a dense
matter description using Skyrme's chiral model~\cite{Skyrme} was
originally suggested by Skyrme and others~\cite{DENSESKYRME}. At
large $N_c$ and low density matter consisting of solitons
crystallizes as the ratio of potential to kinetic energy
$\Gamma=V/K\approx N_c^2/p_F^2\gg 1$ is much larger than 1.
The crystal melts at sufficiently high density with
$\Gamma\approx N_c^2/p_F^2\approx 1$, or sufficiently
high temperature with $\Gamma\approx N_c/T\approx 1$.
QCD matter at large $N_c$ was recently revisited in~\cite{ROBE}.

The many-soliton problem can be simplified in the crystal limit by
first considering all solitons to be the same and second by reducing
the crystal to a single cell with boundary conditions much like the
Wigner-Seitz approximation in the theory of solids. A natural way to
describe the crystal topology is through $T^3$ with periodic boundary
conditions. A much simpler and analytically tractable approximation consists of
treating each Wigner-Seitz cell as $S^3$ with no boundary condition
involved. The result is dense Skyrmion matter on $S^3$ \cite{DENSES3}.

At low baryonic densities holographic QCD is a crystal of instantons with
the Wigner-Seitz cell {\it approximated} by $S^3$. The
pertinent instanton is defined on $S^3\times R$~\cite{KSZ3}. At moderate
densities chiral symmetry is restored on the average with an $n_B^{5/3}$
equation of state~\cite{KSZ3}. This homogenous (on the average) liquid-like
phase is strongly coupled and not emmenable to standard Fermi liquid analysis.

In this paper, we would like to follow up on the transport properties
in the homogeneous phase originally discussed in~\cite{KSZ2} using D4/D8
to contrast them with some recent studies in~\cite{KSS} using D3/D7.
In section 2, we recall the bulk characteristics of the homogeneous phase
in D4/D8 and suggest that it may be  identified with a strongly coupled
holographic liquid prior to the restoration of chiral symmetry. In section 3,
we derive the general formulae for the holographic currents induced by an
external baryonic field in the linear response approximation for both
D4/D8 and D3/D7. In section 4 we show that the transverse
baryonic current for cold D4/D8 is saturated by medium modified vector
mesons in leading $N_c$ in agreement with~\cite{KSZ2}. The bulk static
conductivity is zero. Large $N_c$ D4/D8 is an insulator. In section 5, we
develop the quasi-normal mode approach for hot and dense D4/D8 and D3/D7,
both of which are conductors at large $N_c$.
For completeness we also discuss cold D3/D7 in light of a recent result
\cite{KSS}. In section 6, we suggest a unified visco-elastic framework
for interpreting gapless excitations in dense media in both the elastic
(collisionless) and hydrodynamic (collision) regimes. We argue that
cold D3/D7 exhibits such a mode at large $N_c$
with zero bulk viscosity and finite
shear viscosity. In section 7, we suggest that the leading $1/N_c$
correction to the baryonic currents in cold D4/D8 can be extracted
from an effective baryonic theory using the Random Phase Approximation.
Our conclusions and prospects are in section 8. A number of points
pertaining to transport in dense holographic media are discussed
in the Appendices.

\section{Homogeneous dense matter}

First we briefly review the bulk property of the homogeneous phase in the
SS model. More details can be found in~\cite{KSZ2}. We consider Sakai-Sugimoto's original
embedding \cite{Sakai1, Sakai2},
where the D8-branes configuration in the $\t$ coordinate is constant and not affected by
the existence of a background $U(1)_V$ field $\mathbb{A}_0$.
This corresponds to $\t ={\d\t}/{4}$, the maximal asymptotic separation between D8 and
$\DeB$ branes. The DBI action of D8 branes with $\mathbb{A}_0$ is written as\footnote{The integral
is extended to $(-\infty, \infty)$ to take into account $\DeB$
branes as well as D8 branes.}
\begin{eqnarray}
S_{\mathrm{DBI}} =  -a \int d^4x \int dZ\, K^{2/3}  \, \sqrt {\, 1 - b
K^{1/3}
  (\dell_Z \mathbb{A}_0)^2 } \ ,\label{Action.without.vector}
\end{eqnarray}
where
\begin{eqnarray}
a \equiv \frac{N_c N_f \l^3 \Mkk^4}{3^9 \pi^5} \ ,  \qquad b
\equiv \frac{3^6 \pi^2}{4 \l^2 \Mkk^2} \ , \qquad  K = 1+Z^2 \ .
\end{eqnarray}
$\Mkk$ is the Kaluza-Klein mass and $\l$ is t'Hooft coupling.
Now we introduce the baryon source coupled to $\mathbb{A}_0$ through
the Chern-Simons term. We assume that baryons are uniformly distributed over
$\mathbb{R}^3$ space whose volume is $V$. For large $\lambda$,
the instanton size is $1/\sqrt{\lambda}$ \cite{Baryons1, Baryons2}. It
can be treated as a static delta function source at large $N_c$.
For a uniform baryon distribution, the source is
\be
   S_{\mathrm{source}}= N_c n_B \int d^4x \int dZ\, \delta (Z)\mathbb{A}_0(Z)\ .
\label{SOURCE}
\ee

By varing the total action $S_{\mathrm{DBI}} + S_{\mathrm{source}}$
we get the classical solution $\mathbb{A}_0$:
\begin{eqnarray}
\mathbb{A}_0(Z;n_q) = \dint{0}{Z}dZ \frac{n_q/2}{\sqrt{{(ab)^2K^2}
+ b K^{1/3}n_q^2/4 }}\ , \label{Az.solution}
\end{eqnarray}
which defines the baryon {\it quark} chemical potential  $\m_q$ as
\begin{eqnarray}
\mu_q(n_q) \equiv \lim_{|Z|\ra \infty} \mathbb{A}_0(Z;n_q)  \ . \label{Mu.Q}
\end{eqnarray}
This relation also defines $\m_q$ as a function of $n_q$ and vice
versa. The baryon chemical potential $\m_B$ is
\begin{eqnarray}
\m_B = m_B + N_c \m_q \ , \label{def.chemical}
\end{eqnarray}
where $m_B$ is the baryon rest mass.

The interaction energy density $\e_{\mathrm{int}}$, pressure $P$, grand potential $\Omega$, and
the baryon chemical potential $\m_B$ have been computed in~\cite{KSZ2},
\begin{eqnarray}
\e_{\mathrm{int}} \equiv \frac{\D E}{V} &=&  a \dint{-\infty}{\infty} dZ
K^{2/3} \left[  \sqrt{{1 + \frac{(N_c n_B)^2}{4 a^2 b}K^{-5/3}}}
-1\right] \
, \\
P = -\frac{\Omega}{V} &=&   a \dint{-\infty}{\infty} dZ\,
K^{2/3} \left[1 - \frac{1}{\sqrt {\, 1 + \frac{(N_c n_B)^2}{4 a^2 b} K^{-5/3} }} \right]\ , \label{Pressure} \\
{\m_B} &=& m_B + N_c \dint{-\infty}{\infty}dZ \frac{N_c
n_B/4}{\sqrt{{(ab)^2K^2} + b K^{1/3}(N_c n_B/2)^2 }} \ ,
\end{eqnarray}
where $V$ is the volume and $\Omega$ is understood as a function of $\m_B$ through (\ref{Mu.Q}) and (\ref{def.chemical}). At low densities they translate to
\begin{eqnarray}
\e_{\mathrm{int}}  &\sim&   \frac{27 \pi^3}{2}\frac{N_c}{N_f \l} \inv{\Mkk^2} n_B^2   \ , \\
P = -\frac{\Omega}{V}  &\sim&   \frac{27 \pi^3}{2}\frac{N_c}{N_f \l} \inv{\Mkk^2} n_B^2   \ , \\
\m_B &\sim& m_B +  27 \pi^3\frac{N_c}{N_f \l} \inv{\Mkk^2} n_B \ .
\end{eqnarray}
The baryonic contributions appear through the combination $N_c/\l N_f$.
The large $N_c$ and large $\l$ limit are not compatible in the homogeneous phase.
Compatibility with solitonic physics suggests that the large $N_c$ limit be taken
first followed by the large $\l$ limit, which is also consistent with holography.
This will be assumed throughout, unless specified otherwise.
\begin{figure}[]
  \begin{center}
    \includegraphics[width=9cm]{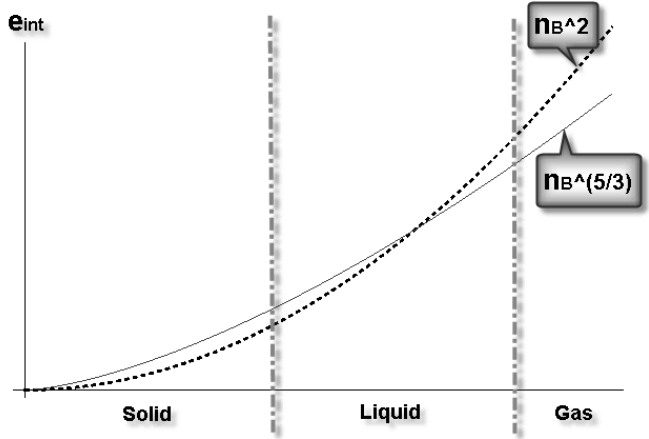}
  \caption{Sketch of the phases of cold D4/D8.}
  \label{Fig:phases}
  \end{center}
\end{figure}
The homogeneous phase described by (\ref{Pressure}) breaks spontaneously chiral symmetry
with density dependent vacuum-like modes \cite{KSZ2}. In Fig.~\ref{Fig:phases} we sketch
the various phases of dense holographic matter at zero temperature. The low density part
is inhomogeneous (solid) with spontaneously broken chiral symmetry, while the high density
phase is homogeneous (gas) with restored chiral symmetry. Intermediate between the two is
a possible liquid phase. Here we suggest that (\ref{Pressure}) may capture some aspects of
the liquid phase still in the spontaneously broken phase using holography. The solid phase
binds with an energy density $\e_{\mathrm{int}} \approx a_M\,n_B$ where $a_M$ is the Madelung
constant for the pertinent crystallization provided that the baryons are semiclassically
quantized to account for the pion-interaction through the mesonic cloud~\cite{KZ}.
The gas phase is homogeneous with $\e_{\mathrm{int}} \approx n_B^{5/3}$
and restored chiral symmetry~\cite{KSZ3}

\subsection{Compressibility}

Holographic QCD at large $N_c$ and large $\l$ is umcompressible.
Indeed, under small scalar or longitudinal vector stress the
baryonic density $n_B$ is expected to change locally to $n_B+\delta n_B$
so that the constitutive equations read
\begin{eqnarray}
&&Mn_B\dot{\vec v}=-\vec\nabla p \ , \label{Newton1} \\
&&\partial_t{\delta n_B}+n_B\vec\nabla\cdot\vec{v}=0 \label{Baryon1} \ ,
\label{CONSTITUTIVE}
\end{eqnarray}
by the Newtonian equation of motion (\ref{Newton1}) and baryon current conservation (\ref{Baryon1}). The baryonic charges
move with an acceleration $(\partial P / \partial n_B)/m_B\approx 1/\l$
which is suppressed at large $\l$ since
$m_B = 8\pi \l N_c$~\cite{Baryons2}. Another way to say this is to note
that (\ref{CONSTITUTIVE}) implies $(\partial_t^2-c_1^2\nabla^2)\delta n_B=0$,
with the speed of the first or thermodynamic sound $c_1=\sqrt{\partial P/\partial n_B/m_B}$.
For the confined D4/D8 configuration
\begin{eqnarray}
c_1=  \left(\frac{27\pi}{8}\frac{n_B}{\l^2 N_f} \int dZ \inv{K}\frac{1}{
      \sqrt{1+\frac{N_c^2 n_B^2}{4 a^2 b}K^{-5/3}} ^3} \right)^{1/2} \ ,
\end{eqnarray}
after using (\ref{Pressure}). The bulk modulus is
$\mathbf{K}=n_B\partial P /\partial n_B \approx n_B^2 (N_c/\lambda)$,
with the compressibility $\chi=1/\mathbf{K}\approx (\l /N_c)/n_B^2$.
Holographic QCD is uncompressible at large $N_c$.

\section{Holographic Baryonic Currents}

\begin{figure}[]
  \begin{center}
    \includegraphics[width=14cm]{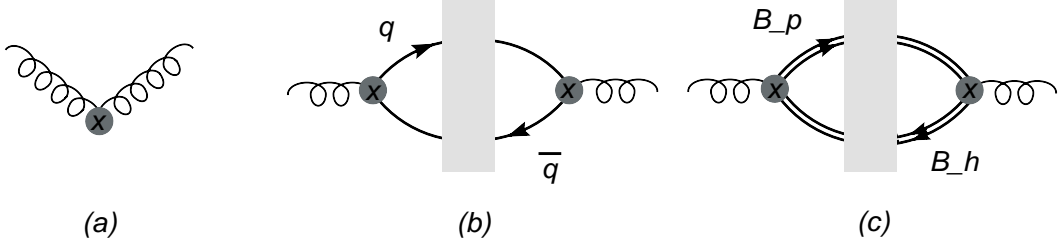}
  \caption{Typical contributions to the baryonic response in D4/D8.
           (a) Direct $N_c^0$ (b) Vector mesons $N_c^0$ and (c) Fermi (baryon) contributions $N_c^{-1}$}
  \label{Fig:feynman0}
  \end{center}
\end{figure}

Baryon transport in confined D4/D8 occurs explicitly through $1/N_c$ effects.
Contributions to the baryonic current to order $N_c^0$ are shown in Fig.\ref{Fig:feynman0}. They follow
from direct (a) or vector meson (b) such as the $\w$ meson. All density effects
in holography are suppressed at large $N_c$ and large $\l$. To illustrate
these points, we streamline the dense analysis given in~\cite{KSZ2} using
general notations to extend the results to finite temperature and also
other brane embeddings.
The induced metric on the D8 branes for both low (KK) and high temperatures(BH)
can be written as
\begin{eqnarray}
ds_{\mathrm{D8}}^2 &\equiv& g_{tt} dt^2 +g_{xx}
\d_{ij}dx^{i}dx^{j} + {g_{UU}}\, dU^2 + g_{SS}
d\Omega_4^2 \  \label{D8metricabs} \\
&\equiv&  \a\left(\frac{U}{R}\right)^{3/2} dt^2 +
\left(\frac{U}{R}\right)^{3/2}\d_{ij}dx^{i}dx^{j} +
\left(\frac{R}{U}\right)^{3/2} \g\, dU^2 +
\left(\frac{R}{U}\right)^{3/2}U^2d\Omega_4^2 \ , \label{D8metric}
\end{eqnarray}
where for the KK background
\begin{eqnarray}
\a \ \ra\ -1 \  , \quad \g \ \ra\ \inv{f(U)} + \left(\frac{\dell
x^4}{\dell U}\right)^2 \left(\frac{U}{R} \right)^3 f(U) \ , \quad
f(U)\ \ra\ 1-\left(\frac{\Ukk}{U}\right)^3 \ ,  \label{KK}
\end{eqnarray}
and for the BH background
\begin{eqnarray}
\a \ra -f(U) \  , \quad \g \ra \inv{f(U)} + \left(\frac{\dell
x^4}{\dell U}\right)^2 \left(\frac{U}{R} \right)^3 \ , \quad
f(U)\ \ra\ 1-\left(\frac{U_T}{U}\right)^3 \ . \label{BH}
\end{eqnarray}
The embedding information is only encoded in $\gamma$ and thereby
$g_{UU}$.

With the induced metric (\ref{D8metricabs}) and the pertinent
guage fields, the general DBI action follows as
\begin{eqnarray}
  && S_{\mathrm{DBI}} = - \caln\ \Tr \int d^4x dU \ e^{-\phi} g_{SS}^2
      \left[ - g_{00}g_{xx}^3g_{UU} - g_{xx}^3 F_{0U}F_{0U}
      - g_{xx}^2g_{UU} \sum_i F_{0i}F_{0i} \right. \nn \\
  &&  \qquad \qquad \left. - g_{00}g_{xx}^2 \sum_i F_{iU}F_{iU} - g_{00}g_{xx}g_{UU} \sum_{i>j} F_{ij}F_{ij}
      - g_{xx} \sum_{i>j} F_{ij}F_{ij}F_{U0}F_{U0} + \cdots \right]^{1/2}   \label{DBI}
\end{eqnarray}
where $ e^{-\phi} = g_s(U/R)^{3/4}$ and $\caln \equiv T_8 \Omega_4$.
The $D_8$ brane tension is $T_8$ and $\Omega_4$ is the volume of a unit $S^4$.
\footnote{We absorb $2\pi\a'$ into the gauge field for notational convenience.
It will be recalled in the final physical quantities.}
The $F^3$ and $F^5$ terms cancel by symmetry. Among the $F^4$ terms we only
retained the relevant term for our discussion below. If we consider the fluctuation ($A_\a(x^\a)$) around the
classical configuration $\mathbb{A}_0$ (\ref{Az.solution}), which is due to homogeneous matter at $Z=0$, the action can be expanded as
\begin{eqnarray}
  && \frac{R^{2/3}\caln}{2 g_s} (2\pi\a')^2 \Tr \int dU U^{5/2} \inv{\sqrt{-{\a}\g}}
  \left[ \frac{2 \a \g \D^{-1}}{(2\pi\a')^2} + 2 \D (\dell_U \mathbb{A}_0) F_{U0} + \D^3 F_{U0}F_{U0}  \right. \nn  \\
  && \qquad \qquad  \ \left. + \D \a \sum_i F_{iU}F_{iU} + \D \g \left(\frac{R}{U}\right)^3 \sum_i F_{0i}F_{0i}
  +\D^{-1}\a\g\left(\frac{R}{U}\right)^3 \sum_{i>j} F_{ij}F_{ij}
  \right] \ ,
\end{eqnarray}
up to quadratic terms. $F_{\a\b} \equiv \dell_\a A_\b - \dell_\b A_\a - i [A_\a, A_\b]$ and
\begin{eqnarray}
  \D \equiv \frac{1}{\sqrt{1 +  \frac{(2\pi\a')^2}{\a\g}(\mathbb{A}_0')^2}}\ \ .
\end{eqnarray}
It is useful to change variable
\begin{eqnarray}
U = U_0 (1 + Z^2)^{1/3} \ , \label{UtoZ}
\end{eqnarray}
where $U_0$ is the coordinate of the tip of the $\De$-$\DeB$
cigar-shaped configuration in the confining background and the
position of the horizon $U_T$ in the black hole background. The range
of $Z$ is $(0,\infty)$ contrary to $U$ whose range is
$(U_0,\infty)$. Also this range can be extended to
$(-\infty,\infty)$ if we consider $\DeB$ branes $(-\infty,0)$
together with $\De$ branes $(0, \infty)$ in a natural way. For
convenience, we note the following useful relations
\begin{eqnarray}
K \equiv 1+Z^2 \ , \quad U = U_0K^{1/3} \ , \quad dU = \frac{2U_0
}{3}\frac{Z}{K^{2/3}} dZ\ , \quad f =
1-\left(\frac{U_*}{U_0}\right)^3\inv{K} \ ,
\end{eqnarray}
where $U_*$ is $U_T$ for the black hole background. For the
confining background, from here on and for simplicity, we follow
Sakai and Sugimoto~\cite{Sakai1} and choose $U_* = \Ukk$.
In terms of $Z$ the action reads
\begin{eqnarray}
  && S = N \Tr \int d^4x dZ k_1
  \left[ 2 \D \mathbb{A}_0' F_{Z0} + \D^3 F_{Z0}F_{Z0} + \D k_2  \sum_i F_{0i}F_{0i}  \right. \nn  \\
  && \qquad \qquad \qquad \qquad \qquad \quad \left. + \D k_3 \sum_i F_{iZ}F_{iZ}
  +\D^{-1} k_2k_3  \sum_{i>j} F_{ij}F_{ij} \label{GeneralAction}
  \right] \ ,
\end{eqnarray}
where we dropped the fluctuation independent part and the parameters ($k_1, k_2, k_3, \D, N$) are different
for each of the brane embeddings. They are summarized in Table 1. The case $D3/D7$ is separatly discussed below
(section \ref{5.3}).
We note that the dimensionless densities are $d = \frac{3^{6}\pi^4}{\l^2 N_f\Mkk^3} n_B$
for D4/D8$_\mathrm{con}$, $d = \frac{3^5\pi^2}{2^5N_cN_f\l T^5 l_s^2} n_B$ for D4/D8$_\mathrm{dec}$,
and $d = \frac{(2\pi)^3}{\sqrt{\l} N_f N_c} n_q$ for D3/D7 with $f=1-\frac{Z_H^4}{Z^4}$. This explicitly
shows that the density effects are subleading at large $\l$ and large $N_c$ for fixed $N_f$.

\begin{table}[]
\begin{center}
\begin{tabular}{|c|c|c|c|c|c|c|}
\hline
\vspace{-0.4cm}& & & &  & & \\
 &  $N$ & $k_1$ & $k_2$ & $k_3$    & $2\pi\a'\mathbb{A}_0'$ &  $\D$   \\
\vspace{-0.4cm}& & & &  & & \\
\hline \hline\
\vspace{-0.4cm}& & & &   & & \\
 D4/D8$_{\mathrm{con}}$ & $  \k \equiv \frac{\l N_c}{216\pi^3} $ & $  K
$ & $K^{-4/3}\Mkk^{-2} $ &  $-1$       &  $ \frac{d}{\sqrt{b}\sqrt{K^2 +  K^{1/3}d^2}}  $ &
$ \sqrt{1 + d^2 K^{-5/3}
}$
\\
\vspace{-0.4cm}& & & &    & &\\
D4/D8$_\mathrm{dec}$ & $ \frac{\l N_c T^3}{54\pi}  $ & $ \frac{K^{3/2}}{\sqrt{K-1}}  $ & $K^{-4/3}(2\pi T)^{-2}$ & $-\frac{K-1}{K}$
 &      $ \frac{d}{\sqrt{K^{5/3} + d^2}} $  &  $ \sqrt{1+ d^2 K^{-5/3} }$ \\
\vspace{-0.4cm}& & & & & & \\
D3/D7 & $\frac{\l N_f N_c}{2(2\pi)^4} $ & $ Z^3 $ & $Z^{-4} f^{-1} $  & $-f$    & $ \frac{d}{\sqrt{Z^6 + d^2}} $
 &  $\sqrt{1+ d^2 Z^{-6} }$ \\
\hline
\end{tabular}
\end{center}
   \caption{Parameters of the different embeddings in (\ref{GeneralAction}). See text.}
    \label{Table:Parameters}
\end{table}

Now consider an abelian fluctuation in the $ A_Z = 0 $ gauge
\begin{eqnarray}
  && A_\m = a_\m(x^0, x^3 ,Z) + \calv_\m(x^0, x^3) \ ,  \label{Gauge}
\end{eqnarray}
where $a_\m$ vanishes at the boundary i.e. $a_\m(x^0, x^3, \infty)=0$,
so that the boundary field is simply $\calv_\m(x^0, x^3)$.
$\calv_\m(x^0, x^3)$ exists for all $Z$ as the background.
\footnote{This is equivalent to the usual set up $A_\m$ with the boundary condition $A_\m(x^0, x^3 , \infty) = \calv_\m(x^0, x^3) $.
The difference is in the equations of motion. The equation for $A_\m$ is
homogeneous but the equation for $a_\m$ is inhomogeneous and sourced by $\calv_\m$
as shown in (\ref{EOMLT}).}
With the Fourier decomposition
\begin{eqnarray}
  && a_\m(Z,x^0,x^3) = \int \frac{d\w dq}{(2\pi)^2} e^{-i\w x^0 + iqx^3} a_\m
  (Z,\w,q)\ ,  \\
  && \calv_\m(Z,x^0,x^3) = \int \frac{d\w dq}{(2\pi)^2} e^{-i\w x^0 + iqx^3} \calv_\m
  (\w,q)\ ,
\end{eqnarray}
the quadratic action can be  rewritten as
\footnote{For simplicity we omitted the argument of the functions. Each quadratic term
is a function of ($\w,q$) (first) and ($-\w,-q$) (second). We dropped
the surface terms since they vanish on shell when the source is explicitly present.}
\begin{eqnarray}
  && S = N \int \frac{d\w dq}{(2\pi)^2} dZ \Big[ a_L \cald_L a_L  -2 f_L \calv_L a_L - f_L \calv_L \calv_L - 2 g_L a_L    \nn \\
  && \qquad \qquad \qquad \qquad   +  a_T \cald_T a_T  -2 f_T \calv_T a_T - f_T \calv_T \calv_T  \Big]  \ , \nn
\end{eqnarray}
where we introduced the gauge invariant variables
\begin{eqnarray}
  \begin{array}{cllll}
    \mathrm{Longitudinal\ mode}:& & a_L \equiv q a_0 + \w a_3, & &\calv_L \equiv q \calv_0 + \w \calv_3\ , \\
    \mathrm{Transverse\ mode}: & &a_T \equiv \w a_1,  & &V_T \equiv \w \calv_1\ ,
  \end{array}
\end{eqnarray}
with $a_2=0$  and used Gauss constraint $\D^2 \w a_0' + k_3 q a_3' = 0 $.  $a_2=0$ is a consistent choice since the transversal equation of motion decouples  from the others.

The differential operators $\cald_{L/T}$ are defined as
\begin{eqnarray}
  && \cald_L \equiv \dell_Z \frac{- k_1 k_3 \D^3}{\D^2 \w^2  + k_3 q^2 } \dell_Z + k_1 k_2 \D \ , \label{zerolon} \\
  && \cald_T \equiv \inv{\w^2} \Big( \dell_Z k_1 k_3 \D \dell_Z - k_1 k_2(\D \w^2 + \D^{-1}k_3 q^2) \Big) \ , \label{zerotrans}
\end{eqnarray}
and the coefficient functions are
\begin{eqnarray}
  && f_L \equiv - k_1 k_2 \D \ , \qquad  g_L \equiv \frac{k_3 q \mathbb{A}_0'}{ \D^2\w^2 + k_3 q^2 }  \ , \\
  && f_T \equiv \frac{k_1 k_2(\D \w^2 + \D^{-1}k_3 q^2)}{\w^2} \ .
\label{fg}
\end{eqnarray}
The equations of motion is
\begin{eqnarray}
  && \cald_L a_L = f_L \calv_L + g_L  \ , \qquad \cald_T a_T = f_T \calv_T . \label{EOMLT}
\end{eqnarray}
With the formal solutions
\begin{eqnarray}
 && a_L =  \cald_L^{-1} (f_L \calv_L + g_L) \ , \qquad a_T =   \cald_T^{-1} (f_T \calv_T) \ ,
\end{eqnarray}
the on-shell action reads
\begin{eqnarray}
  && S = - N \int \frac{d\w dq}{(2\pi)^2} dZ \bigg( \calv_L f_L [ \cald_L^{-1}(f_L \calv_L)  +\calv_L] +
        \calv_L[2f_L \cald_L^{-1}g_L] + g_L \cald_L^{-1} g_L  \nn \\
        && \qquad \qquad \qquad \qquad \quad +  \calv_T f_T [ \cald_T^{-1}(f_T \calv_T)  +\calv_T] \bigg) \ .
\end{eqnarray}

The induced baryonic currents follow to leading order in large $N_c$ and large $\lambda$ as
\begin{eqnarray}
  && J_L(\w,q) =  2 N \w \int dZ f_L [ \cald_L^{-1}(f_L \calv_L + g_L) + \calv_L] \ , \\
  && J_T(\w,q) =  2 N \w \int dZ f_T [ \cald_T^{-1}(f_T \calv_T) + \calv_T] \ ,
\label{JLJT}
\end{eqnarray}
where $\cald^{-1}_L$ and  $\cald^{-1}_T$ are understood with the retarded prescription
$\omega\rightarrow \omega +i0$. $f_L, f_T, g_L$ are all recorded in (\ref{fg}). The
longitudinal current involves $g_L$ independently of $\calv_L$ as $g_L$ is triggered
by the {\it gradient} of the baryonic profile $\mathbb{A}_0'$. This is the analogue
of Fick's law (baryonic charge diffusion). The terms involving $\calv_{L,T}$ correspond
to $\sigma_{L,T}$ the longitudinal and transverse Fourier transforms of the
space-time conductivities. The arguments $(\w, q)$ are subsumed.

\section{D4/D8: Cold}

In the confined phase, the operators $\cald_{L,T}$ are hermitian modulo the retarde
prescription in frequency space. They can be diagonalized using eigenmodes as
discussed in~\cite{KSZ2}. Throughout the prescription $\omega\rightarrow \omega+i0$
is subsumed.

\subsection{Longitudinal Mode}

\begin{figure}[]
  \begin{center}
    \includegraphics[width=8cm]{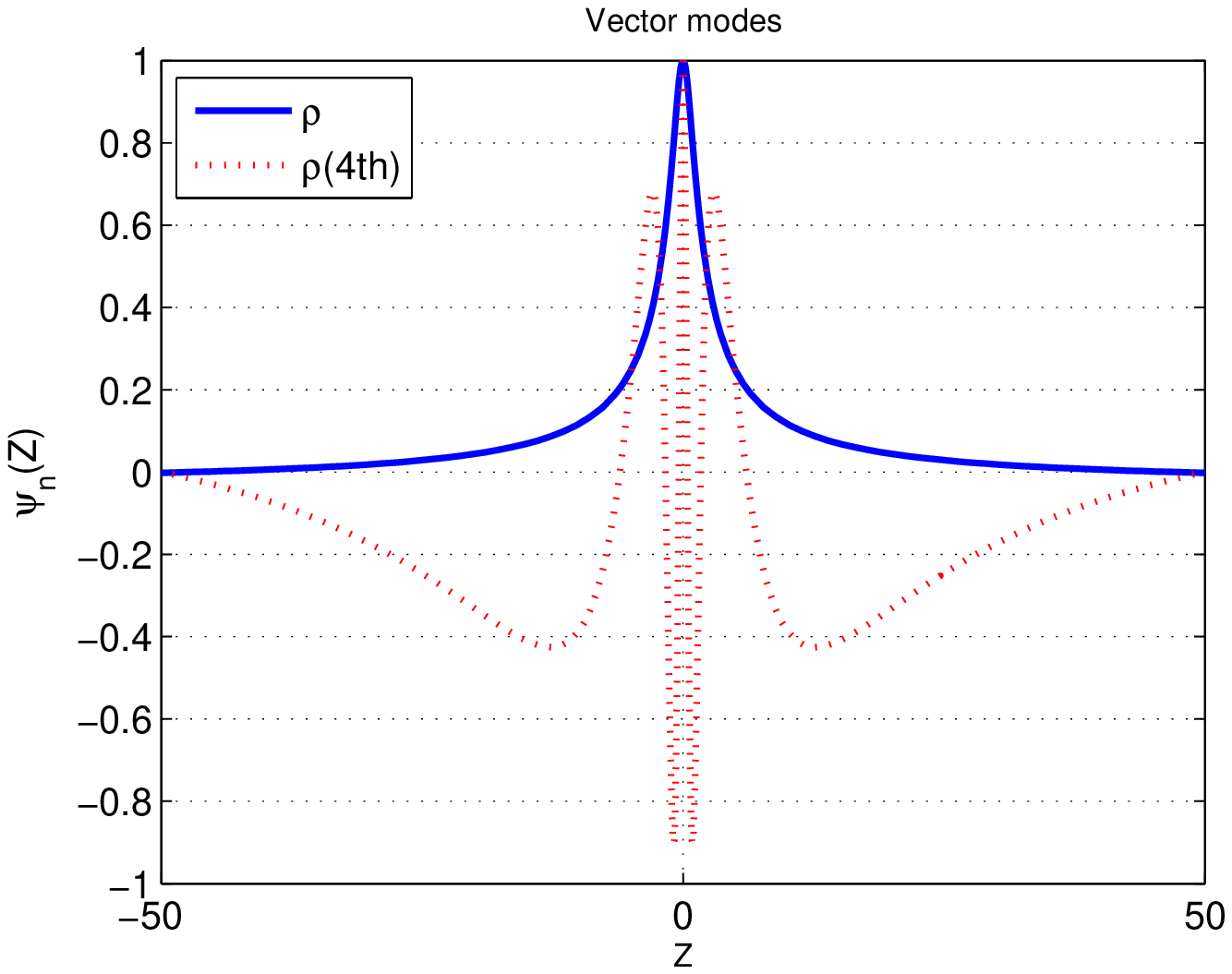}
    \includegraphics[width=8cm]{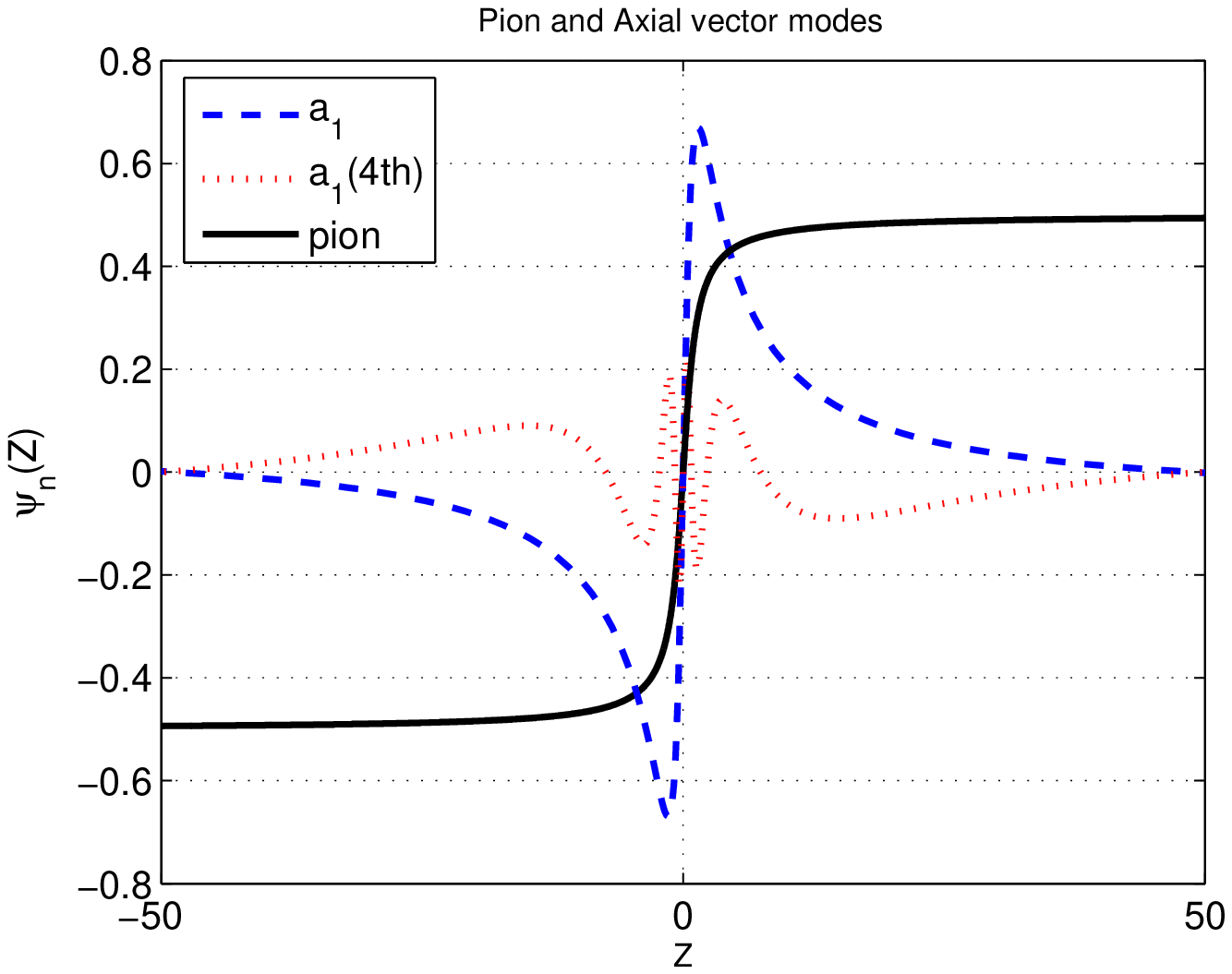}
  \caption{Vector mode functions (Left) and axial-vector and pion mode functions (Right)}
  \label{Fig:modes}
  \end{center}
\end{figure}

The longitudinal operator ($\cald_L$) is
\begin{eqnarray}
  && \cald_L \equiv \dell_Z \frac{ K \D^3}{\D^2 \w^2  -  q^2 } \dell_Z + K^{-1/3} \D \ .
\end{eqnarray}
When $q=0$ or $\w=0$ it is easily diagonalized, since
\begin{eqnarray}
    \cald_L(q=0) &=& \inv{\w^2} \dell_Z  K \D   \dell_Z + K^{-1/3} \D \ , \nn \\
    \cald_L(\w=0) &=& -\inv{q^2} \dell_Z  K \D^3    \dell_Z + K^{-1/3}  \D \ .
\end{eqnarray}
The Green's function ($\cald^{-1}_L$) may be expanded in terms of the complete set of eigenvalues
that diagonalize
\begin{eqnarray}
    && \cald_L(q=0) f  =  \left( \frac{K^{-1/3} \D}{\w^2} \right) \l f\ ,  \nn \\
    && \cald_L(\w=0) f  =  \left( \frac{K^{-1/3} \D}{q^2} \right) \l f\ ,
\end{eqnarray}
where $  \frac{K^{-1/3}\D}{\w^2} $ and  $\frac{K^{-1/3}\D}{q^2}  $ are weight factors. Using
the complete sets,
\begin{eqnarray}
  && (\dell_Z K \D \dell_Z)\, \chi_n = - (K^{-1/3} \D) \,  (\l^{\chi}_n)^2 \, \chi_n \ , \nn \\
  && (\dell_Z K \D^3 \dell_Z)\, \xi_n = - (K^{-1/3} \D) \,  (\l^{\xi}_n)^2 \, \xi_n \ ,
\label{XSET}
\end{eqnarray}
we have
\begin{eqnarray}
  && \bra{Z} \cald_{L}^{-1}(q=0) \ket{Z'} = \sum_{n \in \mathbb{N }} \frac{\chi_n(Z)
\chi_n(Z')}{-\w^2 + (\l^\chi_n)^2} + \frac{\chi_0(Z) \chi_0(Z')}{-\w^2} \ ,\label{Green.w.zero} \\
  && \bra{Z} \cald_{L}^{-1}(\w=0) \ket{Z'} = \sum_{n \in \mathbb{N }} \frac{\xi_n(Z)
\xi_n(Z')}{q^2 + (\l^\xi_n)^2} + \frac{\xi_0(Z) \xi_0(Z')}{q^2} \ , \label{Green.q.zero}
\end{eqnarray}
which are the results of \cite{KSZ2}. Typical behaviors of $\chi_n$ and $\xi_n$ are shown in Fig.\ref{Fig:modes}.

For small $\omega, q$ we may write without loss of generality,
\begin{eqnarray}
   \bra{Z} \cald_{L}^{-1}(\w \approx q \approx 0) \ket{Z'}
   &\approx&  \sum_{n \in 2 \mathbb{N}} \frac{ \varphi_n(Z) \varphi_n(Z')}
   {- \w^2 + c_n^2 q^2  + (\l^\chi_n)^2 } + \frac{\varphi_0(Z) \varphi_0(Z')}{-\w^2 + c_{\pi}^2 q^2} \nn \\
   &+& \sum_{n \in 2 \mathbb{N}+1 } \frac{ \varphi_n(Z) \varphi_n(Z')}
   {- \w^2 + c_n^2 q^2  + (\l^\chi_n)^2 } \  . \label{Green.w.q}
\end{eqnarray}
The first contribution is from the density dependent axial-vector mode, the second
contribution is from the density dependent pion mode
(strictly speaking its U(1) partner at large $N_c$), and the last contribution is from
the density dependent vector mode. The denominators are the dispersive modes, while the
numerators capture their residues. The even-odd in the labelling of the modes translates
into odd-even in the parity of $\varphi_n(Z)$. The baryonic current reads
\begin{eqnarray}
  && J_L =  2N \w \int dZ K^{-1/3} \D \left[ \cald^{-1}_L (K^{-1/3} \D \calv_L + g_L) + \calv_L\right] \ .
\end{eqnarray}
The Z-integration picks only the vector or even-modes of (\ref{Green.w.zero}) since
$\calv_L$ is trivially even. The longitudinal baryonic conductivity in the confined
case is
\begin{eqnarray}
\sigma_L=2N \w \int dZ K^{-1/3} \D \left[ 1+\cald^{-1}_L K^{-1/3} \D  \right]\,\,.
\end{eqnarray}
The longitudinal mesonic propagator $\cald^{-1}_L$ admits the mode decomposition
(\ref{Green.w.q}). From (\ref{XSET}) it follows that
\begin{eqnarray}
\int dZ f_L\chi_n=\frac 1{\l_n^2}\int dZ\,(\dell_Z K \D \dell_Z)\, \chi_n =
\frac 1{\l_n^2}\,\left( K \D \dell_Z\, \chi_n\right)^{+\infty}_{-\infty}=0\nn
\label{ZERO}
\end{eqnarray}
is a zero boundary term. The longitudinal baryonic conductivity simplifies
\begin{eqnarray}
\sigma_L=-2N\w\int dZ f_L \ ,
\end{eqnarray}
and so does the longitudinal current. The longitudinal conductivity vanishes
at $\w=0$. Confined holographic QCD is a static insulator at large
$N_c$ and large $\l$ in agreement with our recent analysis~\cite{KSZ4}.


We now note that
\begin{eqnarray}
  c_n \equiv \frac{\l^\chi_n}{\l^\xi_n} \ ,
  \qquad c_\pi \equiv \frac{f_\pi^S}{f_\pi^T} = \sqrt{\frac{\int dZ K^{-1} \D^{-3}}{\int dZ K^{-1} \D^{-1}}} ,
  \label{pionvelocity}
\end{eqnarray}
where $f_\pi^S$ and $f_\pi^T$ have been derived in \cite{KSZ2}. At high density the pion
speed vanishes as $ c_\pi\approx 1/n_B$. The propagation of the axial charge
stalls in very dense matter. For small momenta $q$ the poles develop at
\begin{eqnarray}
  \w_n \approx \sqrt{(c_n^2 q^2 +  \l^\chi_n)^2 } \approx \l^\chi_n  +\half \frac{c_n^2 q^2}{\l^\chi_n} \ ,
\end{eqnarray}
while for small frequencies $\w$
\begin{eqnarray}
   c_n^2(- (\l^\xi_n)^2) +  (\l^\chi_n)^2 = 0 \  \Rightarrow c_n^2 = \left(\frac{\l^\chi_n}{\l^\xi_n}\right)^2 \ ,
\end{eqnarray}
since $q_n^2 = - (\l^\xi_n)^2$ from (\ref{Green.q.zero}).  In the confined D4/D8 embedding,
the vector and axial modes disperse through
\begin{eqnarray}
  \w_n \approx \l^\chi_n  +\half \frac{\l^\chi_n}{(\l^\xi_n)^2} q^2 \ ,
\end{eqnarray}
\begin{figure}[]
  \begin{center}
    \includegraphics[width=8cm]{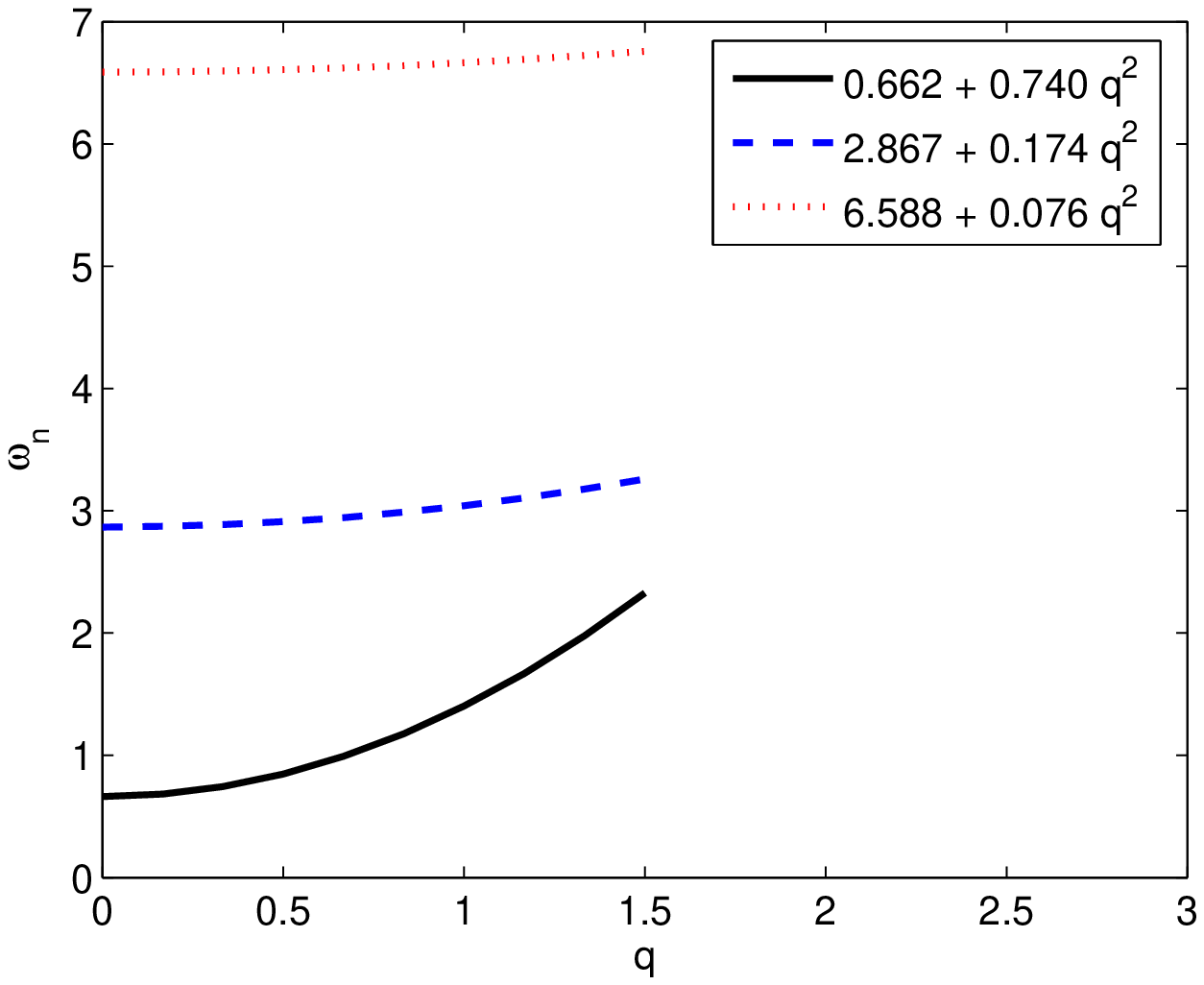}
    \includegraphics[width=8cm]{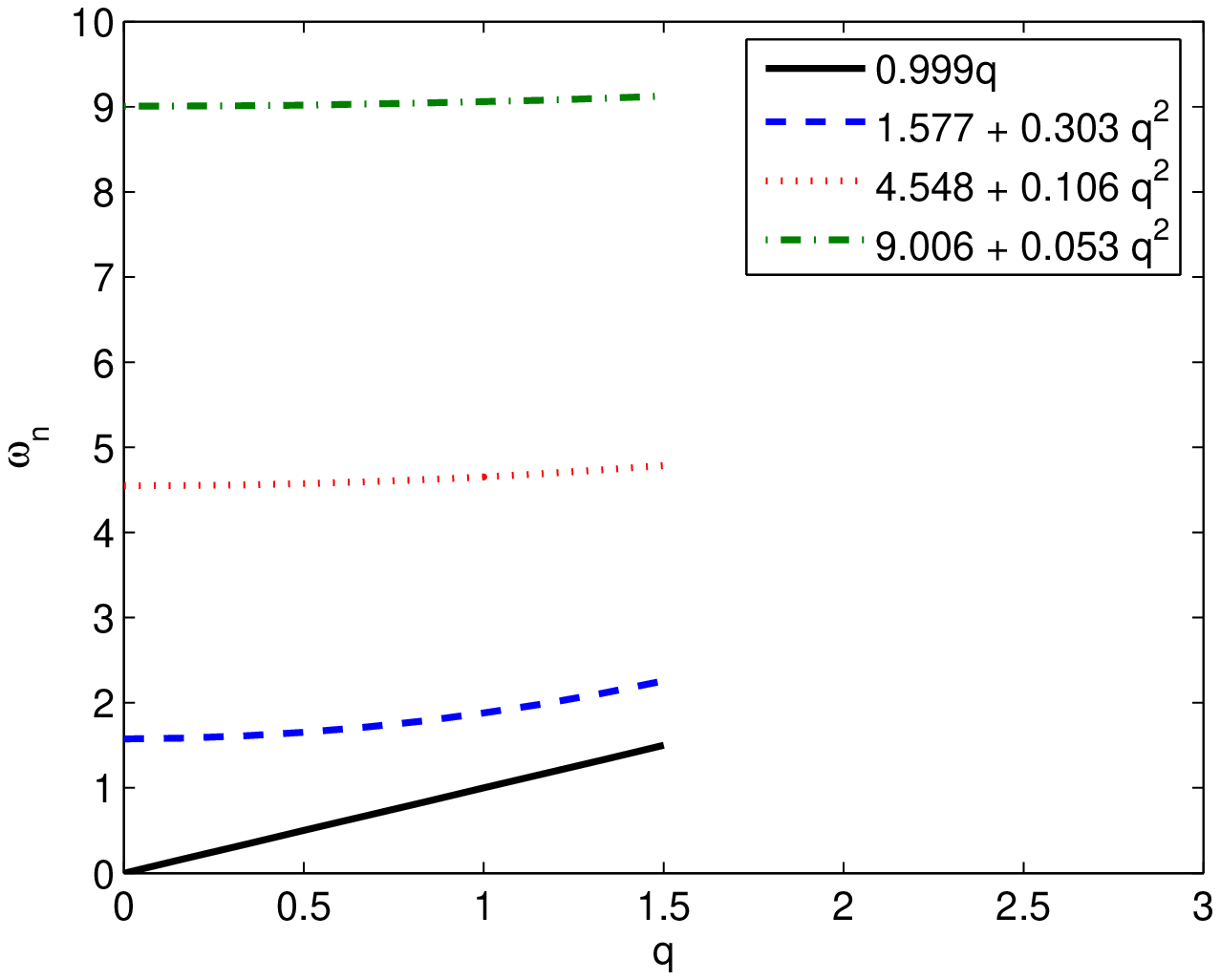}
  \caption{Dispersion relation for vectors (Left) and  axial-vectors including the massless pion (Right), $n=1.26 n_0$}
  \label{Fig:Dispersion}
  \end{center}
\end{figure}
where $\l^\chi_n$ is the rest mass and $\frac{(\l^\xi_n)^2}{\l^\chi_n}$ is the kinetic mass (Fig.\ref{Fig:Dispersion}).
To this order, the imaginary parts vanish in holographic QCD~\cite{KSZ2}. Indeed, vector, axial-vector and pionic
modes are expected to be absorbed by excited and/or recoiling baryons which are $1/N_c$ suppressed effects in cold and
dense QCD.

\subsection{Transverse Mode}

The transverse operator ($\cald_T$) is given by
\begin{eqnarray}
  && \cald_T \equiv -\inv{\w^2} \left(  \dell_Z K \D \dell_Z + K^{-1/3}(\D \w^2 - \D^{-1} q^2) \right) \ .
\end{eqnarray}
For $q=0$  it diagonalizes trivially through
\begin{eqnarray}
    \cald_T (q = 0) \equiv -\inv{\w^2} \dell_Z K \D \dell_Z - K^{-1/3} \D  \ ,
\end{eqnarray}
The Green's function ($\cald^{-1}_L$) may be expanded in terms of the complete set of eigenvalues
that diagonalize
\begin{eqnarray}
    && \cald_T(q=0) f  =  - \left( \frac{K^{-1/3} \D}{\w^2} \right) \l f\ ,
\end{eqnarray}
by using the eigenvalues
\begin{eqnarray}
  (\dell_Z K \D \dell_Z)\, \chi_n = - (K^{-1/3} \D) \,  (\l^{\chi}_n)^2 \, \chi_n \ .
\end{eqnarray}
The Green's function is then expanded as
\begin{eqnarray}
   \bra{Z} \cald_{T}^{-1}(q=0) \ket{Z'} = \sum_{n \in \mathbb{N}} \frac{\chi_n(Z)
\chi_n(Z')}{-\w^2 + (\l^\chi_n)^2} + \frac{\chi_0(Z) \chi_0(Z')}{-\w^2} \ ,
\end{eqnarray}
which is the same as $\cald_{L}^{-1}$.  A rerun of the arguments for the longitudinal current
response yields
\begin{eqnarray}
  && J_T(\w,q) =  2 N \w \int dZ f_T [ \cald_T^{-1}(f_T \calv_T) + \calv_T] \ ,
\end{eqnarray}
where again the retarded prescription is subsumed and
\begin{eqnarray}
  && f_T =\frac{K^{-1/3}(\D \w^2 - \D^{-1} q^2)}{\w^2} \ .
\end{eqnarray}
In the confined phase, the transverse conductivity follows
\begin{eqnarray}
\sigma_T=2 N \w \int dZ f_T [ 1+ \cald_T^{-1} f_T ] \  .
\end{eqnarray}
Using the vector meson mode decomposition for $\cald_T^{-1}$ (\ref{Green.w.q})
and the relation (\ref{ZERO}) we can simplify the transverse conductivity
\begin{eqnarray}
\sigma_T=\sigma_L+2 N \w \int dZ (f_T+f_L) [ 1+ \cald_T^{-1} (f_T+f_L) ] \  ,
\end{eqnarray}
with
\begin{eqnarray}
f_T+f_L=-\frac{q^2}{\w^2}\,K^{-1/3}\Delta^{-1}\,\,.
\end{eqnarray}
The transverse conductivity is vector meson mediated as shown in Fig.\ref{Fig:feynman0}.
For $q=0$, $\sigma_T=\sigma_L$ and vanishes for $\w\rightarrow 0$ in agreement
with~\cite{KSZ4}.

\section{Quasi-normal mode analysis}

we now turn our attention to the deconfined phase of dense holographic
models with non-hermitean or absorptive boundary conditions. For that,
the retarded prescription on the inversion of $\cald_{L,T}$ is best
captured by the quasi-normal mode analysis. The latter is enforced
analytically by matching for the gapless modes and numerically for
the gapped modes. We now present the general formulas
pertinent to the longitudinal and transverse currents.

For $\w \ll 1, q \ll 1$ the equations (\ref{zerolon}) and (\ref{zerotrans}) are reduced to
\begin{eqnarray}
  && \left(\dell_Z \frac{-k_1 k_3 \D^3}{\D^2 \w^2 + k_3 q^2}\dell_Z\right) a_L = 0 \ ,  \label{Start} \\
  && \left(\dell_Z k_1 k_3 \D \dell_Z\right) a_T = 0\ .
\end{eqnarray}
%
%
The general solutions are
\begin{eqnarray}
  && a_L(Z) = C_L \int_Z^\infty dZ \left(\frac{\w^2}{-k_1k_3 \D} + \frac{q^2}{-k_1\D^3}\right) \ ,\label{general_lon} \\
  && a_T(Z) = C_T \int_Z^\infty dZ \left(\frac{1}{-k_1k_3 \D} \right) = a_L(Z)(q\ra0, \w \ra 1) \ , \label{genreal_trans}
\end{eqnarray}
where we imposed the vanishing Dirichlet boundary condition at the boundary $Z=\infty$ i.e. $a_{L/T}(\infty) = 0$.
$C_{L/T}$ will be determined by imposing incoming boundary condition at $Z=0$ which corresponds to
the location of matter (confined phase) or the black hole horizon (deconfined phase).

To constrain $C_{L/T}$  we need to know the behavior of  $a_{L/T}$ around $Z=0$. First, we solve
the equations (\ref{zerolon}) and (\ref{zerotrans}) near $Z=0$ with fixed $\w$ and $q$. Second, we
take the limit $\w \ll 1, q \ll 1$. $\D_L a_L = 0$ and $\D_T a_T = 0$ may be written as
\begin{eqnarray}
  && a_L'' + \left[ \frac{\D' (3k_3q^2 +  \w^2 \D^2(1 - k_3 (1/k_3)' \D/\D'))}{\D(k_3 q^2+\w^2\D^2)}  + \frac{k_1'}{k_1}
  \right]a_L' - k_2\frac{\w^2 \D^2 + k_3 q^2}{  k_3 \D^2} a_L = 0 \ , \\
  && a_T'' + \left[ \frac{(k_1k_3\D)'}{k_1k_3\D} \right] a_T' - k_2 \frac{ \w^2 \D^2 + k_3 q^2}{ k_3 \D^2} a_T = 0  \ .\label{End}
\end{eqnarray}
In this analysis the variable $Z$ introduced in (\ref{UtoZ}) is not convenient due to
the complicated form of $\D$. To make it simpler we introduce the new variable:
\begin{eqnarray}
  z = \frac{1}{(1+Z^2)^{1/3}},
\end{eqnarray}
which is nothing but $U_0/U$ in terms of the original coordinate in (\ref{D8metric}).
In this coordinate the boundary is $u=0$ and the horizon or the matter location is $u=1$.

\subsection{D4/D8: Cold and Dense}

Before analyzing the deconfined phase of D4/D8 with black-hole background absorption,
it is amusing to ask whether the D4/D8 confined background with matter in the KK
background could be also addressed with {\it absorptive} or non-hermitean boundary
conditions. After all {\it cold} matter disperses and absorbs waves much
like a black-hole. From (\ref{general_lon}) and (\ref{genreal_trans}) it
follows that
\begin{eqnarray}
  && a_L =  C_L (\w^2 a_\w(z)- q^2 a_q(z))  \ , \\
  && a_T =  C_T a_\w(z) \ ,
\end{eqnarray}
where
\begin{eqnarray}
  && a_q(z) \equiv \int_0^z dz' \sqrt{ \frac{z'}{(1+d^2z'^5)^3}} \frac{1}{\sqrt{1-z'^3}} \ ,  \\
  && a_\w(z) \equiv \int_0^z dz' \sqrt{\frac{z'}{1+d^2z'^5}}\frac{1}{\sqrt{1-z'^3}} \ ,
\end{eqnarray}
for $\w, q\ll 1$. Recall that the density $d$ is defined by $\frac{3^6\pi^4}{\l^2 N_f \Mkk^3}n_B$ below (\ref{GeneralAction}).
At small densities we expand
\begin{eqnarray}
  a_L(z) = C_L\left[ (\w^2 - q^2) a^{(0)}(z)  - \frac{d^2}{2} (\w^2 - 3q^2) a^{(1)}(z)
                     +\frac{3d^4}{8}(\w^2 - 5q^2) a^{(2)}(z)   \right]
                 + \calo(d^6) \ ,
\end{eqnarray}
with
\begin{eqnarray}
  && a^{(0)}(z) = \int_0^{z} dz' \sqrt{\frac{z'}{1-z'^3}} =  -\frac{2}{3} \left( \arcsin (\sqrt{1 - z^3}) -  \frac{\pi}{2}\right) \ , \nn \\
  && a^{(1)}(z) = \int_0^{z} dz' \sqrt{\frac{z'}{1-z'^3}}\ z'^5 \nn \\
  && \qquad = -\frac{1}{60}\sqrt{1 - z^3} (3 \sqrt{z} (7 + 4 z^3) +
   7\ _2F_1[1/2, 5/6; 3/2; 1-z^3])  + \frac{7\sqrt{\pi}}{120} \frac{\G(1/6)}{ \G(2/3)} \ , \nn \\
  && a^{(2)}(z) = \int_0^{z} dz' \sqrt{\frac{z'}{1-z'^3}}\ z'^{10} \nn \ .
\end{eqnarray}
We don't show $a^{(2)}(z)$ explicitly since it is long and not illuminating.

To impose the {\it incoming boundary condition} we expand the solution around $z=1$
\begin{eqnarray}
  && a_L(z) = C_L A + C_L B \sqrt{1-z} + \calo((1-z)^{3/2}) \ , \label{smallq.con}
\end{eqnarray}
where
\begin{eqnarray}
  && A \equiv  -\frac{\pi}{3}(q^2- \w^2) +
     \frac{ 7 d^2 \sqrt{\pi} (3 q^2 - \w^2) \G(1/6)  }{ 240 \G(2/3)}
     + \frac{187 d^4 \sqrt{\pi} (-5 q^2 + \w^2) \G(5/6)}{ 3584\G(4/3) } \ , \nn \\
  && B \equiv \frac{(-8  +12 d^2 -15 d^4) q^2 + (8 - 4 d^2 + 3 d^4) \w^2}{ 4 \sqrt{3}} \ ,
\end{eqnarray}
On the other hand we may first solve the equation near z=1. In general $\D_L a_L = 0$ and $\D_T a_T = 0$ are
\begin{eqnarray}
  && a_L'' + \left[ \frac{ (5 d^2 z^4)\{3q^2 - \w^2(1+d^2z^5) \} }{2(1 + d^2 z^5)\{ q^2 - \w^2(1+d^2z^5)\}}
      - \frac{3z^2}{2(1-z^3)} - \inv{2z} \right]a_L' \nn \\
  && \qquad \qquad \qquad \qquad\qquad \qquad \qquad \qquad \qquad
  - \frac{9\{q^2 - \w^2(1+d^2z^2)\}}{4z(1-z^3)(1+d^2z^5)} a_L = 0 \ . \\
  && a_T'' + \left[ \frac{ (5 d^2 z^4) }{2(1 + d^2 z^5)}
      - \frac{3z^2}{2(1-z^3)} - \inv{2z} \right]a_L'
  - \frac{9\{q^2 - \w^2(1+d^2z^2)\}}{4z(1-z^3)(1+d^2z^5)} a_L = 0 \ ,
\end{eqnarray}
and reduce to
\begin{eqnarray}
  && a_{L/T}'' - \half \inv{1-z}a_{L/T}' + \frac{3 }{4} \frac{\w^2 - \frac{q^2}{1+d^2}}{(1-z)}a_{L/T} = 0 \ ,
\end{eqnarray}
near $z=1$. The solutions are
\begin{eqnarray}
  && a_{L/T} = C_I e^{- i \sqrt{3\g(1-z)}} \ ,\qquad \g \equiv  \w^2 - \frac{q^2}{1+d^2} \ ,
\end{eqnarray}
where we imposed the incoming boundary condition at $z=1$. Their expanded form reads
\begin{eqnarray}
  && a_L(z) \approx C_I - i C_I \sqrt{3 \g} \sqrt{1-z} + \calo(1-z) \ ,  \label{horizon.con} \\
  && \g = \w^2 - q^2 + q^2d^2 -q^2 d^4 + \calo(d^6) \ .
\end{eqnarray}

A comparison of (\ref{smallq.con}) and (\ref{horizon.con}) yields
\begin{eqnarray}
  B+iA\sqrt{3\g} = 0 \ , \label{constraint.con}
\end{eqnarray}
and the dispersion relation is
\begin{eqnarray}
  \w = \pm \left[1-\frac{d^2}{2} + \frac{3d^4}{8} + \calo(d^6) \right]q + \calo(d^6) q^2 + \calo(d^6) q^3 + \calo(q^4) \ \ .
\end{eqnarray}
The latter resums to
\begin{eqnarray}
  \w = \pm \frac{1}{\sqrt{1+d^2}} q \ .
\end{eqnarray}
This is consistent with the zero mode result obtained in ({\ref{Green.w.q}), (\ref{pionvelocity})
and Fig.\ref{Fig:Dispersion} where we also found the zero mode solution odd in $Z$.
Interestingly enough, the quasinormal mode analysis when applied to the confined
and dense KK background with absorptive boundary condition, it yields a massless pole which is the
pion pole with a speed $c_\pi=1/\sqrt{1+d^2}$. Note that there is {\it no imaginary} part. The
reason can be traced back to the $+$ (outgoing) and $-$ (incoming) wave assignment in
(\ref{constraint.con}), both of which solve
\begin{eqnarray}
   0 = B\pm iA\sqrt{3\g} = \left[\frac{8-4d^2+3d^4}{4\sqrt{3}} \sqrt{\g} \pm i\sqrt{3}\right]\sqrt{\g} = 0 \ ,
\end{eqnarray}
for $\g=0$.

For the transverse mode we may follow the same procedure with the substitution in (\ref{smallq.con})
\begin{eqnarray}
  a_T(z) = a_L(z) (q \ra 0 \ ,  \w \ra 1) \ .
\end{eqnarray}
The relation (\ref{constraint.con}) yields
\begin{eqnarray}
  \w = i \left[ \frac{2}{\pi} + \left( -\frac{1}{\pi} + \frac{7 \G(1/6)}{40 \pi^{3/2} \G(2/3)}
        \right) d^2 + \calo(d^4) \right] + \calo(q) \ ,
\end{eqnarray}
which shows that there is no massless excitation. This channel is indeed gapped in
the confined D4/D8 case as we discussed earlier for the case of reflecting boundary
conditions.

\subsection{D4/D8: Hot and Dense}

The absorptive boundary condition is more appropriate for the deconfined BH
background that we now discuss. For that, we rerun the same steps as we did
in the previous section. First we solve the equations for $\nw \ll 1, \nq \ll 1$,
where $\nw \equiv \frac{\w}{2\pi T}$ and   $\nq \equiv \frac{q}{2\pi T}$.
From (\ref{general_lon}) and (\ref{genreal_trans})
\begin{eqnarray}
  && a_L =  C_L (\nw^2 a_\nw(z)- \nq^2 a_\nq(z))  \ , \\
  && a_T =  C_T a_\nw(z) \ ,
\end{eqnarray}
where
\begin{eqnarray}
   a_\nq(z)  &\equiv&   \int^{z}_0 dz' \sqrt{\frac{z'}{(1+d^2 z'^5)^3}}    \\
   &=& \frac{2}{15} z^{3/2}\left[\frac{3}{\sqrt{1+d^2 z^5}} +  2
  \ _2F_1[3/10, 1/2; 13/10; -d^2 z^5]\right] \ , \nn \\
   a_\nw(z)  &\equiv&   \int^{z}_0 dz' \sqrt{\frac{z'}{1+d^2 z'^5}}\inv{1-z'^3} \ .
\end{eqnarray}
Recall that the density $d$ is defined by $\frac{3^5\pi^2}{2^5\l N_c N_f T^5 l_s^2} n_B$ below (\ref{GeneralAction}).
To impose the incoming boundary condition we expand  $a_L$ around the horizon.
\begin{eqnarray}
  && a_L(1-\e) = a_L(1) - \e a_L'(1) + \cdots \nn \\
  && \qquad    = C_L\nw^2 a_\nw(1)   - \frac{C_L\nw^2}{3\sqrt{1+d^2}} - C_L \nq^2 a_\nq(1) + \calo(\e)  \ , \label{expand.w0}
\end{eqnarray}
where $a_\nw(1)$ has a logarithmic divergence and $a_\nq(1)$ is finite.

In general $\D_L a_L = 0$ and $\D_T a_T = 0$ are
\begin{eqnarray}
  && a_L'' +  \left[ \frac{ (5 d^2 z^4)\{3(1-z^3)q^2 - \nw^2(1+d^2z^5) \} }
     {2(1 + d^2 z^5)\{(1-z^3) \nq^2 - \nw^2(1+d^2z^5)\}} \right. \nn \\
   && \left. \qquad \qquad \qquad \qquad \qquad \qquad   + \frac{\nw^2(1+d^2 z^5)}{(1-z^3)\nq^2
- \nw^2(1+d^2z^5)}\frac{3z^2}{1-z^3} - \inv{2z} \right] a_L' \nn \\
  && \qquad \qquad \qquad \qquad\qquad \qquad
  - \frac{9\{(1-z^3)\nq^2 - \nw^2(1+d^2z^2)\}}{4z(1-z^3)^2(1+d^2z^5)} a_L = 0 \ , \\
  && a_T'' + \left[ \frac{ (5 d^2 z^4) }{2(1 + d^2 z^5)}
      - \frac{3z^2}{(1-z^3)} - \inv{2z} \right] a_T'
  - \frac{9\{(1-z^3)\nq^2 - \nw^2(1+d^2z^2)\}}{4z(1-z^3)^2(1+d^2z^5)}  a_T = 0 \ ,
\end{eqnarray}
which simplify to
\begin{eqnarray}
  && a_{L/T}'' - \inv{1-z}a_{L/T}' + \frac{\nw^2}{4(1-z)^2}a_{L/T} = 0 \ ,
\end{eqnarray}
near the horizon with  $z=1$. The incoming wave solution near the horizon is
\begin{eqnarray}
  && a_{L/T} =  (1-z)^{-i \nw/2} F(z) \ ,
\end{eqnarray}
with $F(z)$ a regular function near $z=1$ or $z=1-\e$. Assuming
$ \nw \ln \e \ll 1$ we have
\begin{eqnarray}
  \e^{-i\nw/2}F(1-\e) &=& F(1-\e) - \frac{i\nw}{2} \ln \e F(1-\e) + \cdots \nn \\
  &=& F(1) -  \frac{i\nw}{2} \ln \e|_{\e \ra 0} F(1) +\calo(\e) \ . \label{expand.h0}
\end{eqnarray}
By comparing the singular part of (\ref{expand.w0}) with (\ref{expand.h0}) we get
\begin{eqnarray}
  C_L = \frac{i3\sqrt{1+d^2}}{2\nw }F(1) \ .
\end{eqnarray}
By comparing the regular part of (\ref{expand.w0}) with (\ref{expand.h0}) we get
the dispersive retation for the longitudinal baryonic waves
\begin{eqnarray}
  1 = -\frac{i\nw}{2} - \frac{i3a_\nq(1)\sqrt{1+d^2}}{2} \frac{\nq^2}{\nw} \ .
\end{eqnarray}
For small $\nw$ and $\nq$ but fixed $\nq^2/\nw$ the dispersion relation is
\begin{eqnarray}
  \w &\approx&  - \frac{i3 \sqrt{1+d^2}}{2} \frac{q^2}{2\pi T} \frac{2}{15} \left[\frac{3}{\sqrt{1+d^2}} +  2
  \ _2F_1[3/10, 1/2; 13/10; -d^2 ]\right] \nn \\
  &\approx& -  i \frac{q^2}{2\pi T} \left(1+\frac{2d^2}{13}-\frac{16d^4}{299} +\cdots  \right)  \quad (\mathrm{For\ small}\ d)\nn \\
  &\approx& -  i \frac{q^2}{2\pi T} \left(  \frac{2\G(1/5)\G(13/10)}{5\G(1/2)} d^{2/5} +  \frac{\G(1/5)\G(13/10)}{5\G(1/2)} d^{-8/5} + \cdots  \right) \  \nn \\
  && \qquad \qquad\qquad\qquad\qquad\qquad\qquad\qquad\qquad\qquad(\mathrm{For\ large}\ d) \ ,
\end{eqnarray}
where both the small and large baryon density limits are displayed explicitly. The longitudinal
diffusion constant is
\begin{eqnarray}
D_L&\approx&   \frac{\sqrt{1+d^2}}{2\pi T} \left[\frac{3/5}{\sqrt{1+d^2}} +  \frac {2}{5}
  \ _2F_1[3/10, 1/2; 13/10; -d^2 ]\right] \ .
\end{eqnarray}
For zero baryon density this is $D=1/2\pi T$. In the deconfined phase of D4/D8 the
baryonic charge diffuses whatever the density. This is expected from baryon number
conservation. The presence of the BH in the deconfined phase overwhelms the Fermi
effects.

A rerun of the analysis for the transverse baryonic current follows the substitution
\begin{eqnarray}
  a_T = a_L (\nq  \ra 0 \ , \nw \ra 1) \ .
\end{eqnarray}
Comparing the singular part of (\ref{expand.w0}) and (\ref{expand.h0}) gives
\begin{eqnarray}
  C_T = \frac{i3 \nw \sqrt{1+d^2}}{2 }F(1) \ ,
\end{eqnarray}
and comparing the regular part of (\ref{expand.w0}) and (\ref{expand.h0}) yields
\begin{eqnarray}
  \nw^3 = i2 \ .
\end{eqnarray}

Thus there is no hydrodynamic pole. The transverse baryonic current in dense
and deconfined D4/D8 is still gapped. It is much like a {\it transverse plasmon}.

\subsection{D3/D7: Hot and Dense} \label{5.3}

For comparison, let us consider in this case the non-chiral and non-confining
embedding with D3/D7 at finite temperature and finite density.
We consider the massless quark embedding where analytic solutions
are available~\cite{KO}. The induced metric becomes simply $AdS_5\times S^3$
independent of the gauge field.
\begin{eqnarray}
ds^2 = \frac{Z^2}{R^2}(-f dt^2 + d\vec{x}^2) +  f^{-1}\frac{R^2}{Z^2} dZ^2 + R^2 d\Omega_3^2 \ ,
\quad f\equiv 1-\frac{Z^4_H}{Z^4} \ , \label{D7metric}
\end{eqnarray}
where $R=4\pi g_s N_c\a'^2$ is the curvature radius. We work in units of $R=1$.
$Z_H=\pi T$ where $T$ is the temperature. SUGRA and SYM quantities will be
tied by $\a' =1/\sqrt{\l}$ with $\l = 4\pi g_s N_c$.
With this metric, we compute the DBI action as
\begin{eqnarray}
  && S_{\mathrm{DBI}} = - \caln\ \Tr \int d^4x dZ \ g_{SS}^{3/2}
      \left[ - g_{00}g_{xx}^3g_{ZZ} - g_{xx}^3 F_{0Z}F_{0Z}
      - g_{xx}^2g_{ZZ} \sum_i F_{0i}F_{0i} \right. \nn \\
  &&  \qquad \qquad \left. - g_{00}g_{xx}^2 \sum_i F_{iZ}F_{iZ} - g_{00}g_{xx}g_{ZZ} \sum_{i>j} F_{ij}F_{ij}
           - g_{xx} \sum_{i>j} F_{ij}F_{ij}F_{Z0}F_{Z0} +  \cdots \right]^{1/2} \   \label{DBId7}
\end{eqnarray}
The result is analogous to the D4/D8 case (\ref{DBI}) with three differences:
1) $\caln = T_7 \Omega_3$;  2) there is no contribution from the dilaton;
3) $g_{SS}^{3/2}$ appears instead of $g_{SS}^{4/2}$, since the compact space is $S^3$ not $S^4$.

To consider finite baryon density (or chemical potential) we set the background vector
$U(1)$ field $\mathbb{A}_0(Z)$ in bulk. Its form follows from minimizing the DBI action
(\ref{DBId7}):
\begin{eqnarray}
  && 2\pi\a'{\mathbb{A}}_0' = \frac{{d}}{\sqrt{Z^6 + {d}^2}} \ .
\end{eqnarray}
We explicitly recalled $2\pi\a'$ and $d \equiv \frac{(2\pi)^3}{\sqrt{\l}N_fN_c}n_q$~\cite{KO,KSS}.

Following the analysis in D4/D8 above,
we now consider mesonic fluctuations around the density background $\mathbb{A}_0$.
In the general form cast in~(\ref{GeneralAction}) the quadratic action reads
\begin{eqnarray}
  && S = N \Tr \int d^4x dZ k_1
  \left[  \D^3 F_{Z0}F_{Z0} + \D k_2  \sum_i F_{0i}F_{0i}  \right. \nn  \\
  && \qquad \qquad \qquad \qquad \qquad \quad \left. + \D k_3 \sum_i F_{iZ}F_{iZ}
  +\D^{-1} k_2k_3  \sum_{i>j} F_{ij}F_{ij}
  \right] \ ,
\end{eqnarray}
where the information of the background field $\mathbb{A}_0$ is encoded in $\D$ and
\begin{eqnarray}
  && N = \frac{\l N_f N_c}{2(2\pi)^4}\ , \qquad  \D=\sqrt{1+d^2 Z^{-6}} \ , \nn \\
  && k_1 = Z^3 \ , \qquad  k^2 = Z^{-4}f^{-1}\ , \qquad k^3 = f^{-1} ,
\end{eqnarray}
as in Table.\ref{Table:Parameters}.

In this general form we can use (\ref{Start})-(\ref{End}).
In terms of the variable $z = \frac{Z_H}{Z}$, $\nw=\frac{\w}{2\pi T}$, $\nq \equiv \frac{q}{2\pi T}$,
and $\nd \equiv \frac{d}{(\pi T)^3}$ we have
\begin{eqnarray}
  && a_L =  C_L (\nw^2 a_\nw(z)- \nq^2 a_\nq(z))  \ , \\
  && a_T =  C_T a_\nw(z) \ ,
\end{eqnarray}
with
\begin{eqnarray}
   a_\nq(z)  &\equiv&   \int^{z}_0 dz' \frac{z'}{\sqrt{1+\nd^2 z'^6}^3}    \\
   &=&  z^2 ( \ _2F_1[3/2, 1/3 ; 4/3, -z^6 d^2]) \ , \nn \\
   a_\nw(z)  &\equiv&   \int^{z}_0 dz' \frac{z'}{\sqrt{1+\nd^2 z'^6}}\inv{1-z'^4} \ .
\end{eqnarray}
Note that the integrand in $a_\nw$ exhibits explicitly the BH horizon
at $z'=1$ in units of temperature. At zero temperature the integrand smoothly
reduces from $  1/(1-z'^4 )$ to 1. However, the BH singularity makes the integral
logarithmically divergent at the horizon. As a result, the zero temperature limit
is singular and will be considered separatly next.
To impose the incoming boundary condition we expand  $a_L$ around the horizon.
\begin{eqnarray}
  && a_L(1-\e) = a_L(1) - \e a_L'(1) + \cdots \nn \\
  && \qquad    = C_L\nw^2 a_\nw(1)   - \frac{C_L\nw^2}{4\sqrt{1+\nd^2}} - C_L \nq^2 a_\nq(1) + \calo(\e)  \ , \label{expand.w}
\end{eqnarray}

Alternatively, the zero mode equation ($\cald_L a_L = 0 $) is
\begin{eqnarray}
   && \dell_z^2 a_L + \left[ \frac{ 3 \nd^2 z^{5}}{(1+\nd^2z^{6})}\left(\frac{3(1-z^4)
\nq^2-\nw^2(1+\nd^2z^{6})}{(1-z^4)\nq^2-\nw^2(1+\nd^2z^{6})}\right) \right. \nn \\
 && \qquad \qquad \qquad \qquad\qquad \qquad\left.+\frac{\nw^2 (1+\nd^2z^{6})}{(1-z^4)\nq^2-\nw^2(1+\nd^2z^{6})}
\frac{4z^3}{1-z^4} - \frac{1}{z} \right]\dell_z a_L \nn \\
  && \qquad\qquad \qquad \qquad \qquad \qquad+
\frac{4}{1-z^4}\left(\frac{\nw^2}{1-z^4}-\frac{\nq^2}{1+\nd^2 z^6}\right) a_L  = 0  \ ,  \\
  && \dell_z^2 a_T + \left[ \frac{ 3 \nd^2 z^{5}}{(1+\nd^2z^{6})} -
\frac{4z^3}{1-z^4}- \frac{1}{z} \right] \dell_z a_T \nn \\
 && \qquad \qquad \qquad \qquad \qquad \qquad + \frac{4}{1-z^4}\left(\frac{\nw^2}{1-z^4}-\frac{\nq^2}{1+\nd^2 z^6}\right) a_T  = 0  \ , \label{aTD7}
\end{eqnarray}
which reduces to
\begin{eqnarray}
  && a_{L/T}'' - \inv{1-z}a_{L/T}' + \frac{\nw^2}{4(1-z)^2}a_{L/T} = 0 \ ,
\end{eqnarray}
near the horizon at $z=1$. The incoming wave solution is of the form
\begin{eqnarray}
  && a_{L/T} =  (1-z)^{-i \nw/2} F(z) \ ,
\end{eqnarray}
with $F(z)$ a regular function near $z=1$. For $ \nw \ln \e \ll 1$ we have
\begin{eqnarray}
  \e^{-i\nw/2}F(1-\e) &=& F(1-\e) - \frac{i\nw}{2} \ln \e F(1-\e) + \cdots \nn \\
  &=& F(1) -  \frac{i\nw}{2} \ln \e|_{\e \ra 0} F(1) +\calo(\e) \ . \label{expand.h}
\end{eqnarray}

A comparison of the singular part of (\ref{expand.w}) and (\ref{expand.h}) yields
\begin{eqnarray}
  C_L = \frac{i2\sqrt{1+\nd^2}}{\nw }F(1) \ ,
\end{eqnarray}
and a comparison of (\ref{expand.w}) and (\ref{expand.h}) yields
\begin{eqnarray}
  1 = -\frac{i\nw}{2} - i2a_\nq(1)\sqrt{1+\nd^2} \frac{\nq^2}{\nw} \ .
\end{eqnarray}
For small $\nw$ and $\nq$ with fixed $\nq^2/\nw$ the dispersion relation follows
\begin{eqnarray}
  \w &\approx&  - i \sqrt{1+ \left(\frac{d}{(\pi T)^3}\right)^2} \frac{q^2}{2\pi T} \ _2F_1[3/2, 1/3 ; 4/3; -d^2] \nn \\
  &\approx& -  i \frac{q^2}{2\pi T}\left(1+\frac{1}{8}\left(\frac{d}{(\pi T)^3}\right)^2-
\frac{1}{112}\left(\frac{d}{(\pi T)^3}\right)^4 +\cdots  \right)
\quad \left({ \mathrm{ For\ small \ }}  \frac{d}{(\pi T)^3} \right) \nn \\
  &\approx& -  i \frac{q^2}{2\pi T}\left( \frac{2\G(7/6)\G(4/3)}{\G(1/2)}
\left(\frac{d}{(\pi T)^3}\right)^{1/3} + \frac{\G(7/6)\G(4/3)}{\G(1/2)}
\left(\frac{d}{(\pi T)^3}\right)^{-5/3} - \cdots  \right) \nn \\
   && \qquad \qquad \qquad \qquad \qquad \qquad \qquad \qquad \qquad \qquad \qquad
     \left({ \mathrm{ For\ Large \ }}  \frac{d}{(\pi T)^3} \right) \ ,
\end{eqnarray}
The longitudinal diffusion constant for hot and dense D3/D7 is
\begin{eqnarray}
D_L &\approx&  \frac 1{2\pi T}\,
\sqrt{1+ \left(\frac{d}{(\pi T)^3}\right)^2} \ _2F_1[3/2, 1/3 ; 4/3; -d^2] \ .
\end{eqnarray}
As mentioned earlier the zero temperature limit is singular owing to the
occurence of the BH pole in the issuing integrals.

To analyze the transverse baryonic current
\footnote{See \cite{FAST}  for related work on the dispersion relation}
in the same limit of small $\w, q$
we follow the same procedure with the substitution in (\ref{expand.w})
\begin{eqnarray}
  a_T = a_L (q  \ra 0 \ , \nw \ra 1) \ .
\end{eqnarray}
A comparison of the singular part of (\ref{expand.w}) and (\ref{expand.h}) yields
\begin{eqnarray}
  C_T = i2 \nw \sqrt{1+\nd^2} F(1) \ ,
\end{eqnarray}
while a comparison of the regular part of (\ref{expand.w}) and (\ref{expand.h}) gives
\begin{eqnarray}
  \nw^3 = i2 \ ,
\end{eqnarray}
which is incompatible with the limits. The transverse baryonic mode in hot and dense
D3/D7 is gapped much like the transverse plasmon in dense matter. This reflects on
the long-range nature of the transverse forces in holography. We will comment further on
this point below.

\subsection{D3/D7: Cold and Dense}

In a recent analysis~\cite{KSS} have reported the occurence of
a {\it zero sound} mode in cold D3/D7. For completeness we now
rederive their results using our general result (\ref{general_lon}).
At zero temperature we set $Z_H=0$ in (\ref{D7metric}) and change the variable
to $z={1}/{Z}$.

For $\w \ll 1$ and $q \ll 1$ (\ref{general_lon}) gives
\begin{eqnarray}
  && a_L =  C_L (\w^2 a_\w(z)- q^2 a_q(z))  \ , \\
  && a_T =  C_T a_\w(z) \ ,
\end{eqnarray}
with
\begin{eqnarray}
   a_q(z)  &\equiv&   \int^{z}_0 dz' \frac{z'}{\sqrt{1+d^2 z'^6}^3}  = \inv{2}z^2 ( \ _2F_1[3/2, 1/3 ; 4/3; -z^6 d^2]) \ , \nn \\
   a_\w(z)  &\equiv&   \int^{z}_0 dz' \frac{z'}{\sqrt{1+d^2 z'^6}} =  \inv{2} z^2 ( \ _2F_1[1/2, 1/3 ; 4/3; -z^6 d^2] ) \ ,
\end{eqnarray}
Near the horizon $a_L$ is expanded as
\begin{eqnarray}
  a_L = C_L A \inv z + C_L B + \calo\left(1/z^2\right)   \ , \label{smallq.d3}
\end{eqnarray}
with
\begin{eqnarray}
  A \equiv \frac{\w^2}{d} \ , \qquad B \equiv  \frac{(q^2 - 3 \w^2) d^{-2/3} \G(1/3)\G(1/6) }{ 18 \G(1/2)} \ .
\end{eqnarray}

On the other hand the zero mode equations $\cald_{L/T} a_{L/T} = 0 $ are

\begin{eqnarray}
  && \dell_z^2 a_L + \left[ \frac{ 3 d^2 z^{5}}{(1+d^2z^{6})}\left(1 + \frac{2q^2}{q^2-\w^2(1+d^2z^{6})}\right)  - \frac{1}{z}
  \right]\dell_z a_L + \left(\w^2-\frac{q^2}{1+d^2 z^6}\right) a_L  = 0 \ , \nn \\
  && \dell_z^2 a_T + \left[ \frac{ 3 d^2 z^{5}}{(1+d^2z^{6})} - \frac{1}{z}
  \right]\dell_z a_T + \left(\w^2-\frac{q^2}{1+d^2 z^6}\right) a_T  = 0 \ .
\end{eqnarray}
Near the horizon the incoming solution is
\begin{eqnarray}
   a_L(z) = C_I \frac{e^{iwz}}{z} \ ,
\end{eqnarray}
and for $ \w z \ll 1$
\begin{eqnarray}
   a_L(z) = \frac{C_I}{z} + i\w C_I \ , \label{horizon.d3}
\end{eqnarray}

A comparison of (\ref{smallq.d3}) with (\ref{horizon.d3}) yields
\begin{eqnarray}
  Ai\w = B \ , \label{constraint.d3}
\end{eqnarray}
which yields the dispersion relation reported in~\cite{KSS}
\begin{eqnarray}
  \w = \pm \frac{q}{\sqrt{3}} - \frac{i q^2}{2 p \m_0} + \calo(q^3) \ ,
\label{MASSLESS}
\end{eqnarray}
for a massless excitation. Holographic D3/D7 at arbitrarily small
temperatures is diffusive. It is not at strictly zero temperature
with the occurence of a long-range collective mode. In the next
section we suggest that this is a visco-elastic mode, and thereby
generalize it to massive quarks. Any amount of temperature
(collisions) destroy the Fermi-surface at large $\lambda$ and
large $N_c$. Indeed, while the temperature effects are of order
$N_c^0$ through the underlying BH metric, the density effects
are $1/\l$ and $1/N_c$ suppressed through the $N_F$ embeddings
either D7 or D8.

Finally and for completeness we note that the transverse baryonic
mode follows also from (\ref{smallq.d3}) with the substitution
\begin{eqnarray}
  a_T = a_L (q \ra 0 \ , \w \ra 1) \ . \label{trans}
\end{eqnarray}
From (\ref{constraint.d3}) we get
\begin{eqnarray}
  \w = i \frac{d^{1/3} \G(1/3) \G(1/6) }{\G(1/2) } \ .
\end{eqnarray}
The transverse baryonic current is gapped in cold D3/D7 much like
the current is plasmon-gapped in a metal.

\section{Visco-Elastic Analysis}

The occurence of a collective mode in cold D3/D7 suggests that
collectivity through the possible occurence of a Fermi surface
at strong coupling maybe at work. To understand that, we propose
to understand this collectivity by unifying the hydrodynamical
or collision regime with the elastic or collisionless regime.

In Fig.~\ref{QUASIB} we show different propagating domains for
a wave of frequency $\omega$ and momentum $q$ in liquids. The dashed curves
are typical wave dispersions. The free particle-hole continuum
occupies the lower quadrant. $\tau$ is a typical relaxation time
to equilibrium, say $\tau\approx 1/T$ (hot) and $\tau\approx 1/\mu$
(cold) for conformal and strongly coupled theories. For waves with
$\omega\tau\gg 1$ and large velocities compared to the Fermi velocity
$v_F$ of a quasiparticle, we expect {\it collisionless} wave propagation or
{\it elastic} regime. For waves with $\omega\tau\ll 1$ but still large
velocities compared to the Fermi velocity $v_F$ we expect {\it collision}
wave propagation or {\it hydrodynamic} regime. Typical cold media
behave elastically at low temperature and hydrodynamically at
higher temperature. Thus, an elastic mode can be turned inelastic
by just raising the temperature. This is typically what happens
in liquid $He^3$ where the zero sound transmutes to the first
or thermodynamic sound by changing its frequency or temperature
to interpolate between the collisionless and collision regimes.
To understand these regimes we now introduce a unified
visco-elastic framework.
\begin{figure}[]
  \begin{center}
    \includegraphics[width=8cm]{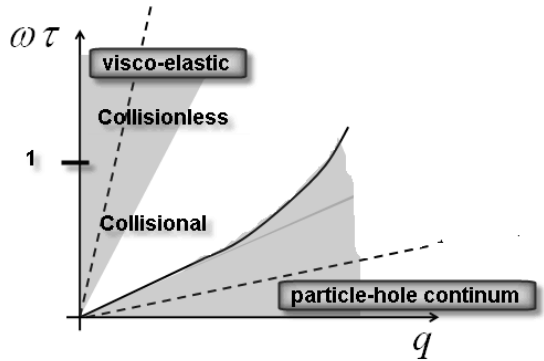}
  \caption{Visco-elastic domain versus the free particle-hole continuum. See text.}
  \label{QUASIB}
  \end{center}
\end{figure}

\subsection{Generalities}

In a homogeneous and isotropic solid, the constitutive equations
for the elastic displacement field $\vec{u} (t,\vec{x})$ are
discussed in the canonical book by Landau and Lifshitz on the
theory of elasticity~\cite{LANDAU}. Specifically
\begin{eqnarray}
m_B\,n_B\frac{\partial^2\vec u}{\partial t^2}=
\left(K+(1-2/p)\,\mu\right)\vec\nabla(\vec\nabla\cdot\vec{u})
+\mu\nabla^2\vec{u}+\vec{F}(t,\vec x)\,\,,
\label{C1}
\end{eqnarray}
where $K$ and $\mu$ are the bulk and shear moduli, $p=3$ is the
dimensionality of space, $\vec{F}$ is an external volume force,
$n_B$ is the baryon equilibrium density and $m_B$ is the bare baryon
mass. The massless case more pertinent for D3/D7 will be deduced
by inspection below. Without loss of generality, we may write
\begin{eqnarray}
\vec{F}(t,\vec x)=n_B\frac{\partial\vec{A}(t,\vec x)}{\partial t}\,\,,
\label{C2}
\end{eqnarray}
which is the 'baryon electric field'. Since the transverse part
of $\vec{A}$ induces a 'baryon magnetic field' we expect
(\ref{C2}) to also include the magnetic contribution as it plays
the role of the Lorentz force. Since we are
interested in the induced baryon current through $\vec{A}$,
the magnetic effects are second order and will be omitted. The
baryon current density is
\begin{eqnarray}
\vec{j}(t,\vec x)=n_B\frac{\partial\vec{u}(t,\vec x)}{\partial t}\,\,.
\label{C3}
\end{eqnarray}

Inserting (\ref{C2}) and (\ref{C3}) in (\ref{C1}) and taking the Fourier transforms
yield
\begin{eqnarray}
-i\omega\,m_B\vec{j}(\omega , \vec{q})=&&
\left(\frac K{n_B} +\left(1-\frac 2p\right)\frac{\mu}{n_B}\right)
\frac{\vec{q}\,(\vec{q}\cdot \vec{j}(\omega, \vec q))}{i\omega}\nn\\
&&+\frac{\mu}{n_B}\frac{q^2}{i\omega}\,\vec{j}(\omega, \vec q)
  -i\omega\,n_B\vec{A}(\omega, \vec q)\,\,.
\label{C4}
\end{eqnarray}
Decomposing the current $j=j_T+j_L$ and the potential
${A}=A_L+A_T$ along $\vec{q}$ and transverse to $\vec{q}$ allows
the transverse and longitudinal induced currents
\begin{eqnarray}
\vec{j}_T(\omega, \vec{q})=
\frac{n_B/m_B}{1-\frac{\mu}{n_B^2}\frac{n_Bq^2}{m_B\omega^2}}\,\vec{A}_T(\omega, \vec q) \ ,
\label{JT}
\end{eqnarray}
and
\begin{eqnarray}
\vec{j}_L(\omega, \vec{q})=
\frac{n_B/m_B}{1-\left(\frac K{n_B^2}+ 2\left(1-\frac 1p\right)\frac{\mu}{n_B^2}\right)
\frac{n_B q^2}{m_B\omega^2}}\,\vec{A}_L(\omega, \vec q)\,\,.
\label{JL}
\end{eqnarray}
$j_L$ relates directly to the induced baryon density through the
local conservation law as $q {j}_L/\omega$.

The response currents (\ref{JT}) and (\ref{JL}) have a direct analogy with their counterparts
in a liquid. Indeed, using hydrodynamics for the baryon current density in a
liquid we can write the analogue of (\ref{C4}). In the linear response approximation
\begin{eqnarray}
-i\omega\,m_B\vec{j}(\omega , \vec{q})=&&
\left(\frac K{n_B} -\frac{i\omega\zeta}{n_B}+\left(1-\frac 2p\right)\frac{i\omega\eta}{n_B}\right)
\frac{\vec{q}\,(\vec{q}\cdot \vec{j}(\omega, \vec q))}{i\omega}\nn\\
&&+\frac{i\omega\eta}{n_B}\frac{q^2}{i\omega}\,\vec{j}(\omega, \vec q)
  -i\omega\,n_B\vec{A}(\omega, \vec q)\,\ ,
\label{C5}
\end{eqnarray}
where the hydrostatic pressure term ${\bf p}$ in the Euler
equation was traded with the longitudinal baryon current through
the continuity equation,
\begin{eqnarray}
-\vec{\nabla}{\bf p}(\omega , \vec q)=-\frac{\partial {\bf p}}{\partial n}\nabla n_B(\omega, \vec q)
=\frac{\partial {\bf p}}{\partial n}\vec{q}\left(\frac{\vec{q}\cdot\vec{j}(\omega, \vec{q})}{i\omega}\right) \ ,
\end{eqnarray}
where $\eta$ and $\zeta$ are the shear and bulk viscosities; ${\bf p}$ is the equilibrium pressure
as a function of density with $K=n_B\partial{\bf p}/\partial n$ the bulk modulus. (\ref{C4}) is very similar
to (\ref{C5}) except for: 1/ the shear modulus in the solid becomes purely imaginary or
$-i\omega\eta$ in the liquid; 2/ the bulk modulus acquires an imaginary part through
$-i\omega\zeta$ in the liquid. Both imaginary parts vanish at $\omega=0$ making the
liquid insensitive to shear at zero frequency. This also means that their contributions
in $j_{L,T}$ are diffusive.

The solid and liquid visco-elastic coefficients can be described in a unified manner through
(\ref{JT}) and (\ref{JL}) by substituting
\begin{eqnarray}
K\rightarrow {\bf K}(\omega) = K(\omega)-i\omega\zeta(\omega) \ ,\\
\mu\rightarrow {\bf M}(\omega) =\mu(\omega)-i\omega\eta(\omega) \ ,
\label{C6}
\end{eqnarray}
as the complex and frequency dependent bulk ${\bf K}$ and shear
${\bf M}$ visco-elastic coefficients. The solid has real ${\bf
M}$ with a small imaginary part $\eta$ (zero up to the uncertainty
principle) while the liquid has imaginary ${\bf M}$ with a small
$\mu$ and large imaginary part $\eta$. The bulk modulus $K$ is
about the same in liquid and solid, and of the same order of
magnitude as the shear viscosity $\eta$.

In light of (\ref{C6}) it follows from (\ref{JL}) and (\ref{JT}) that
the longitudinal current admits a gapless pole (compression mode)
at
\begin{eqnarray}
\omega\approx \left(\frac K {m_Bn_B}+2\left(1-\frac 1p\right)\frac \mu{m_Bn_B}\right)^{1/2}q
-i\left(\frac \zeta{m_Bn_B}+2\left(1-\frac 1p\right)\frac {\eta}{m_Bn_B}\right)\frac {q^2}2 \ ,
\label{FIRST}
\end{eqnarray}
while the transverse current admits a gapless pole (shear mode)
at
\begin{eqnarray}
\omega\approx \sqrt{\frac{\mu}{m_Bn_B}}q-i\frac {\eta}{m_Bn_B}\frac {q^2}2 \ .
\label{ZERO}
\end{eqnarray}
We see that for finite frequency waves and in the long-wavelength
approximation the way a solid responds to external wave-stress is
similar to the way a liquid does. The difference is that in a solid
$\mu\approx K$ and $\zeta, \eta$ are small, while in a liquid
$\mu$ is close to zero and $\zeta, \eta$ are large. In the liquid the
transverse or shear mode becomes diffusive.

\subsection{Cold D3/D7}

D3/D7 at finite density yields a gapped transverse baryonic current
and a gapless longitudinal baryonic current~\cite{KSS}. The gapless
longitudinal baryonic current can be compared with the longitudinal
visco-elastic mode (\ref{FIRST}) with
\begin{eqnarray}
&&\left(\frac K {m_Bn_B}+2\left(1-\frac 1p\right)\frac \mu{m_Bn_B}\right)^{1/2} \Leftrightarrow \frac{1}{\sqrt{p}}
\label{comp1} \ ,\\
&&\left(\frac \zeta{m_Bn_B}+2\left(1-\frac 1p\right)\frac {\eta}{m_Bn_B}\right) \Leftrightarrow \frac 1{p\mu_B} \ . \label{comp2}
\end{eqnarray}
The compressibility is readily tied with the equation of state for any embedding
\begin{eqnarray}
  \frac{K}{m_Bn_B}\equiv \left(\frac{\dell P}{ \dell \e}\right)_S = \frac{1}{{p}} \ ,
\end{eqnarray}
since $\epsilon-pP=0$ in a conformal theory.
\footnote{$p$ is the dimensionality of space and $P$ is pressure.} The energy momentum tensor is still traceless
at finite temperature and density for massless fermions. The gapless and longitudinal
baryonic mode has the speed of the first sound $c_1=1/\sqrt{p}$ for zero shear modulus
$\mu=0$ and massless quarks. For massive quarks $\epsilon-pP\neq 0$. For D3/D7 it follows
from~\cite{KO} that
\begin{eqnarray}
  && \e = \frac{1}{4} \g N(2\pi\a')^4 (\m_q^2-m_q^2)(3\m_q^2 + m_q^2)\ , \\
  && P = \frac{1}{4} \g N(2\pi\a')^4 (\m_q^2-m_q^2)^2 \ ,
\end{eqnarray}
where $\g \approx 0.363 $. The visco-elastic mode has a speed
\begin{eqnarray}
c_1= \sqrt{\left(\frac{\dell P}{\dell \e}\right)_S} = \sqrt{\frac{\m_q^2 - m_q^2}{3\m_q^2 - m_q^2}} \ ,
\label{XSOUND}
\end{eqnarray}
in agreement with the detailed quasi-normal mode analysis in~\cite{KP}.

Since the mode $\omega =q/\sqrt{3}$ lies within the free particle-hole continuum
as shown in Fig.\ref{Fig:feynman0}, it is susceptible to Landau-like damping through single
particle-hole or multi-particle-multi-hole. Again, the visco-elastic analysis
suggests that the bulk viscosity in cold D3/D7 with massless quarks is
\begin{eqnarray}
  \frac{\eta}{n_B} = \frac{\hbar}{2(p -1)} \ , \label{etanb}
\end{eqnarray}
where we used (\ref{chemical}) and (\ref{energy}). In conformal
theories the shear modulus ($\mu=0$) and bulk viscosity are zero ($\zeta$).
For $p=3$, we have $\eta/n_B=\hbar/4$, where $\hbar$ has been restored. This is to
be compared with $\hbar/6\pi$ argued in \cite{GSZ} using the
Stokes-Einstein relation and the uncertainty principle for cold
and strongly coupled Fermions.

It is worth noting  that (\ref{etanb}) can be rewritten as
\begin{eqnarray}
  \frac{\eta}{n_B} = \left(\frac{p}{p-1}\right)\frac{\hbar}{2p}
   = \frac{n_F k_F}{N_c N_f n_B}\frac{\hbar}{p} \ ,
\end{eqnarray}
where
\begin{eqnarray}
  \frac{n_F}{n_B} = \frac{N_c N_f \int_0^{k_F}\frac{d^pk}{(2\pi)^p}\inv{2k} }{ \int_0^{k_F}\frac{d^pk}{(2\pi)^p}}
  = \frac{N_c N_f}{2k_F}\frac{p}{p-1} \ ,
\label{ratio}
\end{eqnarray}
is the ratio of the quark density at the Fermi surface normalized to the baryon density.
For massless quarks $n_B k_F= \e+P$ with  $n_q = N_c N_f n_B $ so that in general
\begin{eqnarray}
 \frac{\eta}{\e + P} = \frac{n_F}{n_q}\frac{\hbar}{p} \ .
\end{eqnarray}
This result can generalized to finite mass by noting that at zero
temperature $\epsilon +P=n_B\mu_B$ and substituting $2k\rightarrow 2\sqrt{k^2+m_q^2}$
in $n_F$ in (\ref{ratio}). Specifically,
\begin{eqnarray}
  && \frac{n_F}{n_B} = \frac{N_c N_f \int_0^{k_F}\frac{d^pk}{(2\pi)^p}\inv{2\sqrt{k^2+m_q^2}} }{ \int_0^{k_F}\frac{d^pk}{(2\pi)^p}}
  = {\frac{N_c N_f v_F}{4k_F}} \, \ _2F_1(1/2, p/2;1+p/2;-v_F^2) \ ,
\end{eqnarray}
with $v_F=k_F/m_q$. Thus
\begin{eqnarray}
\frac{\eta}{n_B}=\left(\frac{\mu_B v_F}{4k_F}\right)\, \ _2F_1(1/2, p/2;1+p/2;-v_F^2)\,\,\frac{\hbar}p\,\,.
\end{eqnarray}
The visco-elastic (\ref{MASSLESS}) for massless quarks,
turn to
\begin{eqnarray}
\omega=c_1q\,-\frac{i}p \left(1-\frac 1p\right)\,
\left(\frac{v_F}{k_F}\,\ _2F_1(1/2, p/2; 1+p/2; -v_F^2)\right)\,\frac{q^2}4\,\,,
\end{eqnarray}
for massive quarks. $c_1$ is the first sound speed. For $D3/D7$ it is explicitly given in (\ref{XSOUND}),
while for arbitrary $Dp/Dq$ it follows from the known equations of state~\cite{KO}.

In D3/D7 any infinitesimal temperature washes out the Fermi surface,
resulting in a diffusive baryonic phase. Thermal collisions at strong coupling
take over the collisions through the Fermi surface however small is the temperature.
As noted earlier, this is the hallmark of strong coupling holography whereby the BH contribution is
of order $N_c^0$ while the Fermi contributions are $1/N_c$ suppressed.

%
%
%

\section{Random Phase Approximation}

If cold D3/D7 exhibits collectivity at large $\l$ and large $N_c$
that is consistent with a constitutive visco-elastic analysis, why
the cold D4/D8 results above are all gapped. The short answer is
that in D4/D8 the baryons are solitons, so baryonic motion with
a Pauli-blocked Fermi surface is subleading in $1/N_c$ as shown
in Fig.~\ref{Fig:feynman0}. Baryons move semi-classically by quantizing the
isorotations and rotations both of which are $1/N_c$ suppressed.
To estimate some of these contributions we note that the baryons
in D4/D8 are flavored instantons with core sizes of order $1/\sqrt{\l}$.
They are heavy with $m_B=8\pi^2N_c\l$. So the semiclassical descriptive
of the translational zero modes follow from the point-like
effective action
\begin{eqnarray}
S=\phi^+\,\left(i\partial_t-\frac{(-i\vec\nabla-{\vec{\cal V}})^2}{2m_B}-\mu_B\right)\,\phi-
\frac 12 \alpha(\phi^+\phi)^2 + .... \ ,
\label{SS}
\end{eqnarray}
where the repulsive interaction $\alpha=(24\pi^4/4M_{KK}^2)(N_c/\l)$ is set to
reproduce the energy density and pressure of the holographic matter (\ref{Pressure})
with $p\approx \alpha n_B^2/2$. The dotted contributions involve higher derivative
terms, e.g. $(\phi^+ (-i\nabla-{\cal V})/m_B\phi)^2$ which are suppressed by $1/N_c$.
Again, note that the limit $N_c$ and $\l$ large do not commute for $\alpha$.
For fixed $\l$ and large $N_c$ the repulsion is strong as it should, while for
fixed $N_c$ and large $\l$ the repulsion is weak. Here $\vec{\cal{V}}$ is the probing
baryonic vector source. In large $N_c$ the baryons are scalar fermions with no assigned
spin to leading order in $1/N_c$. Therefore their spin degeneracy is 1.

In the RPA approximation, the zero modes in (\ref{SS})
integrate to the effective action
\begin{eqnarray}
S_{RPA} ({\cal{V}})= \frac 12 {\cal V}_L\Delta_L{\cal V}_L+\frac 12 {\cal V}_T\Delta_T{\cal V}_T \ ,
\end{eqnarray}
with $j_L=\Delta_L{\cal{V}}_L$ and $j_T=\Delta_T{\cal{V}}_T$ for the longitudinal
and transverse currents. $\w\Delta_{L,T}$ are the longitudinal and transverse baryonic
conductivities respectively. The RPA contributions as shown in Fig.\ref{Fig:fig1}
resum to
\begin{eqnarray}
&&\Delta_L=\frac{\Pi_L}{1-\alpha\frac {q^2}{\omega^2}\Pi_L} \ ,\nn \\
&&\Delta_T=\frac{\Pi_T}{1-{\tilde\alpha}\frac {q^2}{\omega^2}\Pi_T} \ ,
\label{RPA}
\end{eqnarray}
with
\begin{eqnarray}
\Pi_{L,T}=\sum_k^F\frac 1m_B +\sum_k^F\left(\frac {k_{L,T}}{m_B}\right)^2
\Delta_F(k+q)\Delta_F(k) \ .
\label{PI}
\end{eqnarray}
The solid lines in Fig.\ref{Fig:fig1} lie in the Fermi surface. $\tilde\alpha/\alpha\approx 1/N_c$.
The transverse contributions follow solely from the dotted terms in
(\ref{SS}). The summation is carried over the Fermi surface.
$\Delta_F$ is the massive and non-relativistic fermion
propagator associated to (\ref{SS}) in the presence of a Fermi surface.
The first contribution in (\ref{PI}) is from the seagull term in (\ref{SS})
and the second contribution is from the particle-hole bubble.
\begin{figure}[]
  \begin{center}
    \includegraphics[width=11cm]{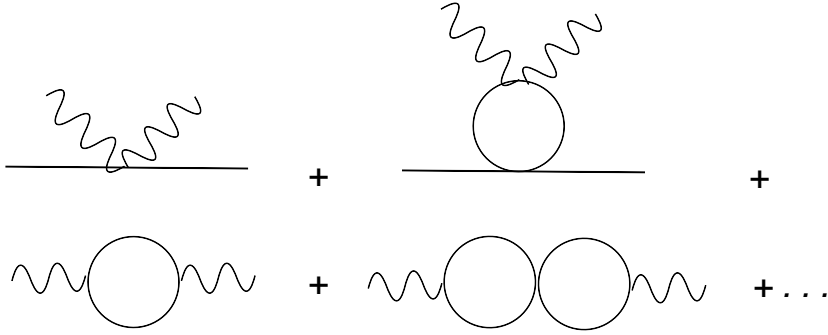}
  \caption{RPA contributions following from (\ref{SS}).}
  \label{Fig:fig1}
  \end{center}
\end{figure}
The londitudinal vector response $\Delta_L$
relates to the scalar density-density response function by current conservation.
Specifically, $\Pi_L=\omega^2/q^2\Pi$ where
\begin{eqnarray}
\Pi(q)=\sum_k^F\Delta_F(k+q)\Delta_F(k)
\label{LIND}
\end{eqnarray}
is the Lindhard function for scalar fermions.
For $\omega, q\rightarrow 0$ but fixed $\beta=\omega/q/v_F$
it takes the form
\begin{eqnarray}
\Pi(q)\approx \frac {m_Bk_F}{2\pi^2}\left(-1+\frac{\beta}2\,{\rm ln}
\left|\frac {1+\beta}{1-\beta}\right|\right)-i\theta(1-|\beta|)\frac{m_B\beta k_F}{4\pi} \ .
\end{eqnarray}
The quasiparticle spectrum following from (\ref{LIND}) is schematically
displayed in Fig.~\ref{QUASIA} with massless quasiparticles of
energy $\omega=v_F q$ at small $k$, and free massive fermions
with $\omega=q^2/2m_B$ away from the Fermi surface.
This description is rooted in weak coupling.
\begin{figure}[]
  \begin{center}
    \includegraphics[width=8cm]{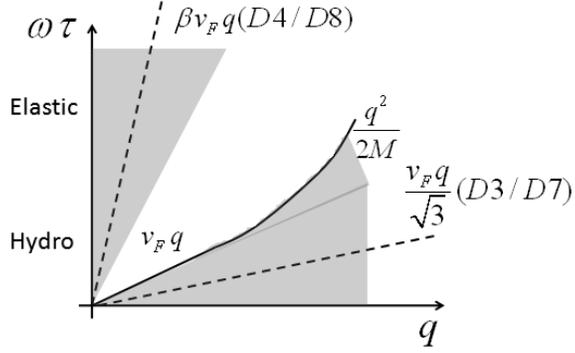}
  \caption{Dispersion relations (dotted lines) for D4/D8 and D3/D7.}
  \label{QUASIA}
  \end{center}
\end{figure}

We note that the longitudinal current in (\ref{RPA})
develop a {\it massless} poles for strongly repulsive fermions
as D4/D8. The longitudinal modes are directly tied to the Lindhard function by
current conservation. They follow from $1=\alpha\Pi$. In particular
the longitudinal sound modes is stable for $\beta>1$
above the quasiparticle cut (no imaginary part) with a speed ($\beta\gg 1$)
\begin{eqnarray}
c_L=\beta\, v_F\approx \sqrt{\frac{\alpha m_Bk_F}{3\pi^2}}\,v_F=\sqrt{\alpha n_B/m_B}=c_1 \ ,
\end{eqnarray}
in agreement with the pressure
$p\approx \alpha n_B^2/2$ in leading order in the density.
For $\beta>1$ the decay of the longitudinal baryonic in
D4/D8 is not through Landau-like damping.

The transverse
mode follows from the pole at $\omega^2=\tilde\alpha q^2\Pi_T$.
For $\tilde\alpha\approx \alpha_*/(N_cq^2)$ the transverse mode is gapped
since
\begin{eqnarray}
\Pi_T(q)\approx \frac {n_B}{m_B} +\frac {n_B}{20m_B\beta^2}-i\theta (1-|\beta|) \frac {3\pi n_B}{8m_B} \ ,
\end{eqnarray}
for $\w , q\rightarrow 0$ and fixed $\beta=\w/q/v_F$. The transverse
gap is typically $\omega_T\approx \alpha_*n_B/m_BN_c$ with $\alpha_*$
following from a a numerical analysis of the transverse quasi-normal
modes in holography. In light of the RPA analysis,
the holographic result {\it suggests} that the transverse
baryonic fluctuations are long ranged and
unscreened unlike their longitudinal counterparts.

\section{Conclusions}

We have analyzed the baryonic transport in D4/D8 (chiral) and D3/D7
(nonchiral) at finite density and/or temperature. D4/D8 is a holographic
model of QCD at large $N_c$ and large 'tHooft coupling $\l$. The
transverse baryonic current in D4/D8 is saturated by the
medium modified vector mesons in the confined phase with $T<M_{KK}/\pi$.
The vector spectrum is gapped since matter is uncompressible at large
$N_c$. Confined D4/D8 matter becomes compressible to order $1/N_c$ with the
occurence of a gapless longitudinal vector mode. While $1/N_c$ effects are
difficult to assess in holography, we have provided an RPA argument for
the speed of the gapless mode using an effective action for baryons
constrained by holography. D4/D8 is diffusive in the deconfined regime.

D3/D7 is diffusive at all temperatures except zero where it is visco-elastic.
This is a hallmark of holography at large $N_c$ and large $\l$. Indeed,
the temperature effects are mediated by the BH background and leading
in $1/N_c$, while the baryonic density effects are carried by the probe
branes which are $N_f/N_c$ suppressed. At strong coupling and large $N_c$
the thermal or collisional collision regime is dominant. The exception is D3/D7 at
zero temperature but finite density as recently pointed by~\cite{KSS}.
A Fermi surface (albeit strongly coupled) maybe  at work in this case
that suggests a visco-elastic regime. A longitudinal gapless mode
emerges with a small width suggestive of a shear viscosity to baryon
ratio $\eta/n_B=\hbar/4$ in cold but dense $D3/D7$.
This mode is turned diffusive by arbitrarily small temperatures
at strong coupling. Our observations extend readily to massive quarks
in D3/D7.

\section{Acknowledgments}
We would like to thank S.~J. Sin for discussions. This work
was supported in part by US-DOE grants DE-FG02-88ER40388 and
DE-FG03-97ER4014.

\appendix

\section{Cold Dp/Dq}

It is interesting to analyze the equation of state of
cold Dp/Dq embeddings. We consider Dq probe branes whose worldvolume spans an $AdS_{p+2}$ factor
and wraps the n-sphere $S^n$ in $AdS_5 \times S^5$, where $p=q-n-1$.
For example, $q=5, p=2$ corresponds to $D5$ branes on $AdS_4 \times S^2$, and
$q=3, p=1$ corresponds to $D3$ branes on $AdS_3 \times S^1$ in $AdS_5 \times S^5$.

At zero temperature and for massless
quarks,  the pressure $P$ and energy density $\e$ read
\cite{KSS}
\begin{eqnarray}
  && P = -\frac{\Omega}{V_p} = \frac{1}{p+1}\frac{\a}{\caln_q^{1/p}} \widetilde{n}_B^{\frac{p+1}{p}}  \ , \\
  && \widetilde{\m}_B =  \frac{\a}{\caln_q^{1/p}} \widetilde{n}_B^{\inv{p}}  \ , \label{chemical} \\
  && \e = -P + \widetilde{\m}_B \widetilde{n}_B =  \frac{p}{p+1}\frac{\a}{\caln_q^{1/p}} \widetilde{n}_B^{\frac{p+1}{p}} \label{energy}\  , \\
  && \frac{\e}{\e_0} = \frac{\a}{4 \pi^{3/2}}\left(\frac{1}{\G(p/2+1)\pi N_cN_f}\right)^{1/p}
    \left(\frac{\l^{\frac{p+1}{2}}}{\caln_q} \right)^{1/p} \ ,   \\
  && \e_0 = 2\sqrt{\pi} \left(\frac{ \G(p/2)p}{4N_cN_f}\right)^{1/p} \frac{p}{p+1}n_q^{(p+1)/p} \ ,
\end{eqnarray}
where $\widetilde{n}_B \equiv \frac{\sqrt{\l}}{2\pi}n_q$, $\wt{\m}_B \equiv \frac{2\pi}{\sqrt{\l}}\m_q $, and $\a \equiv  \frac{\G(1/2 - 1/2p) \G(1+1/2p)}{\G(1/2)}$. $\caln_q \equiv N_f T_{Dq} V_n$ with $V_n$
the volume of a unit n-sphere and $T_{Dq}$  is the $Dq$ brane tension~\cite{KO}. That is $N_7 = \l N_f N_c/(2\pi)^4 $,
$N_5 = \frac{N_f N_c \sqrt{\l}}{2\pi^3} $, and $N_3 = \frac{N_f N_c }{\pi} $. The ratio $\e/\e_0$ shows the energy
density in cold Dp/Dq normalized to the free Fermi energy density:
\begin{eqnarray}
 \frac{\e}{\e_0}&\approx& \l^{1/p} \ .
\end{eqnarray}
For $Dp/Dq \equiv (D3/D7),(D2/D5),(D1/D3)$ dense $Dp/Dq$ is unbound at large $\l$.

\section{Fermionic Drag}

\subsection{Hot D3/D7}

Hot D3/D7 is diffusive at all temperatures. The drag coefficient $\eta_D$
is inversely proportional to the baryonic diffusion constant $D_q$
through the Einstein formulae at strong coupling~\cite{GSZ,MANY}
\begin{eqnarray}
  \eta_D  = \frac{T}{D_q} \ , \label{Deta}
\end{eqnarray}
and for slowly moving particles. In equilibrium, the diffusion constant ties
with the baryonic conductivity $\sigma_q$ (at zero frequency and momentum)
\begin{eqnarray}
\Xi=\frac{<(\Delta N)^2>}{TV_3}=\frac {\sigma_q}{D_q} \ ,
\end{eqnarray}
where $\Xi$ is the baryonic susceptibility. Thus
\begin{eqnarray}
\eta_D=\frac{\Xi}{\sigma_q}T \ .
\end{eqnarray}

The conductivity $\s_q$ has been obtained using Ohm's law in~\cite{KO1}
(See also next Appendix)
\begin{eqnarray}
  \s_q = \frac{N_c N_f T}{4\pi} \sqrt{c^6 + \nd^2} \ ,
   \quad \nd \equiv \frac{8 n_q}{\sqrt{\l} N_c N_f T^3} \ , \label{conductivity}
\end{eqnarray}
with $c\equiv\cos^6\theta(z_*)$ and $z_* \equiv 4/(\pi^2 T^2)$ a
dynamically generated value of a scalar profile at the BH horizon.
Massless quarks correspond to $\theta=0$ and infinite mass quarks to $\theta=\pi/2$.
Thus
\begin{eqnarray}
\eta_D=\frac{\Xi}{N_cN_f}\frac {4\pi}{\sqrt{c^6+\nd^2}} \ .
\end{eqnarray}
This expression expresses the drag of a quark in diffusive D3/D7
for arbitrary mass, temperature and baryon density. When $m_q = 0$
\begin{eqnarray}
\eta_D=\frac{\Xi}{N_cN_f}\frac {4\pi}{\sqrt{1+\nd^2}} = \frac{2\pi T^2}{\sqrt{1+\nd^2}} \ ,
\end{eqnarray}
where $\Xi = \frac{N_fN_cT^2}{2}$~\cite{MST}. When $m_q = \infty$
\begin{eqnarray}
\eta_D=\frac{\Xi}{N_cN_f}\frac {4\pi}{\nd} =\frac{\Xi}{n_q}\,\frac{\pi}2\sqrt{\l}T^3\ ,
\end{eqnarray}
where $\Xi$ can be read off from Eq.(5.7) in \cite{MST}.

\subsection{Cold D3/D7}

The zero temperature case is visco-elastic as we suggested earlier. In this
regime the fermionic conductivity ties to the diffusion constant by the Kubo
formulae
\begin{eqnarray}
  &&   \s_q = n_F D_q \ , \label{sD} \\
  &&   n_F = N_c N_f \int \frac{d^p k}{(2\pi)^p}\inv{2E_k} ( f(E_k) + \bar{f}(E_k)) \ ,
\end{eqnarray}
where $f$ is Boltzmann distribution function and $E_k=\sqrt{k^2+m_q^2}$.
This relation follows from the relaxation time approximation in the quark
probe phase space irrespective of strong or weak coupling. Relaxation to
equilibrium at strong coupling is subsumed. For infinitesimal temperatures
and for finite quark mass~\cite{KO1,MS} (see also next Appendix)
\begin{eqnarray}
  \frac {\s_q}{T} = \frac{N_c N_f}{4\pi}\sqrt{c^6+{\bf d}^2} \ .
\end{eqnarray}
From (\ref{Deta}) it follows that the drag is
\begin{eqnarray}
  \frac{\eta_D}{n_B} = \frac{n_F}{N_c N_f n_B}\frac{4\pi}{\sqrt{c^6 + {\nd}^2}}\ . \label{etanb1}
\end{eqnarray}
This is the general form of the quark drag in a cold holographic
medium (Coulomb phase) with {\it infinitesimal} temperature. We will
assume it also for $T=0$ by continuity.

For $m_q$ finite $(c\ne 0)$ and $N_c, \l \ra \infty$ (${\nd} \ra 0$)
\begin{eqnarray}
  \eta_D \approx  \frac{n_F}{N_c N_f }\frac{4\pi}{c^3} \ .
\end{eqnarray}
When $T=0$ and $m_q=0(c=1)$
\begin{eqnarray}
\eta_D = \frac{2^{(2-p)} \pi^{(1-p/2)}}{\G(p/2)} \frac{\m_q^{p-1}}{p-1} \ ,
\end{eqnarray}
where we used
\begin{eqnarray}
n_F = N_cN_f\int_0^{\m_q} \frac{d^p k}{(2\pi)^p} \inv{2k} \ .
\end{eqnarray}
When $T=0$ and $m_q\ne0(c\ne1)$ and $p=3$
\begin{eqnarray}
  \frac{\eta_D(m_q)}{\eta_D({m_q=0})} = \left[\frac{\m_q}{\sqrt{\m_q^2 - m_q^2}} - \frac{m_q^2}{\m_q^2 - m_q^2}
  \ln \left( \frac{1+ \frac{\m_q}{\sqrt{\m_q^2 - m_q^2}}}{\frac{m}{\sqrt{\m_q^2 - m_q^2}}} \right)\right]\inv{c^3} \ ,
\end{eqnarray}
where $n_q = \sqrt{m_q^2 + k_F^2}$ .

For $m_q \rightarrow \infty (c=0)$
\begin{eqnarray}
  \eta_D \approx  \frac{n_F}{N_c N_f }\frac{4\pi}{{\nd}}
   = \frac{n_F}{n_q} \frac{\pi\sqrt{\l}}{2}T^3 = \inv{2(p-2)}\pi\sqrt{\l}T^2 \ ,
\end{eqnarray}
where $n_q = N_c N_f n_B $ and
\begin{eqnarray}
  \frac{n_F}{n_q} = \frac{\int d^p  k \inv{k^2/2m_q}   e^{-k^2/2m_qT}}{\int d^p k   e^{-k^2/2m_qT} } = \frac{1}{p-2}\inv{T} \ .
\end{eqnarray}
For $p=3$, $\eta_D$ is the drag coefficient reported in~\cite{MANY}.

\section{Baryonic Conductivity}

The baryonic conductivity $\s_q$ in D3/D7 has been derived by various methods~\cite{HK,KO1,MS}.
Generically, the Kubo formulae for the conductivity is
\begin{eqnarray}
 \s_q     = -\lim_{\w \ra 0} \frac{1}{\w} \mathrm{Im} G^{\mathrm{ret}}_{xx}(K) \Big|_{\w = |\vec{k}|} \ ,
\end{eqnarray}
where only the transverse response function contributes, as the longitudinal part
vanishes for light-like momenta by charge conservation. For a rotationally symmetric
medium, $G_{xx}=G_{yy}=G_{zz}$ are the components of the $j_xj_x$, $j_yj_y$ and $j_zj_z$
retarded baryonic current correlations.

Using AdS/CFT the transverse response can be extracted from (\ref{aTD7}), i.e.
\begin{equation}
a_T '' -  \frac{u(u + \nd^2 u(-3+7u^2))}{2(1-u^2)(1+\nd^2u^3)}
a_T ' + \frac{\nw^2 u}{(1-u^2)^2}\frac{1+u\nd^2}{1+u^3\nd^2}\,
a_T =0 \ ,  \label{transv}
\end{equation}
with $u \equiv z^2  $.
The horizon($u=1$) is a regular singular point and the solution behaves as $a_T \sim (1-u)^{\pm i \nw/2}$.
We choose the incoming boundary condition($a_T \sim (1-u)^{- i \nw/2}$) and extract the singularity at $u=1$ by
substituting
\begin{equation}
a_T =(1-u^2)^{-i\nw /2}F(u)\, .
\end{equation}
$F(u)$ is regular at $u=1$ and satisfies the following equation
\begin{eqnarray}
&& F''+\frac{u(-4 + u \nd^2 (3-7u^2)) + 2i(1+u)(1+\nd^2 u^3)\nw}
{2(1-u^2)(1+\nd^2 u^3)} F' \nn \\
&& + \frac{i(1+u)(2+\nd^2u^2(3+5u))\nw + (-1+u + \nd^2u^2(4+3u+u^2))\nw^2} {4(1-u^2)(1+u)(1+\nd^2 u^3)} F =
0\, .
\end{eqnarray}

In the hydrodynamic region($\nw \ll 1$) we may expand $F(u)$ in terms of $\nw$ as
\begin{equation}
F=F_0+\nw F_1+ \nw^2 F_2 \cdots\, ,
\end{equation}
and unwind $F_0, F_1, F_2, \cdots$ order by order.
For that consider the case with $\nd = 0$.
The zeroth order equation is solved with
\begin{equation}
F_0' = \frac{C}{1-u^2} \ ,
\end{equation}
where C is an integration constant. Regularity at $u=1$ forces $C=0$. So $F_0$ is a constant.
At next order we have

\begin{equation}
F_1' = \frac{i}{2(-1+u^2)} F_0 + \frac{C}{-1+u^2}\ .
\end{equation}
Again regularity at $u=1$ sets the constant $C = -iF_0/2$. Thus
\begin{equation}
F_1' = \frac{iF_0}{2(1+u)}\ ,
\end{equation}
which is enough unwinding of the transverse solution for the Green's
function in the zero frequency limit needed for the Kubo formulae.

To obtain the retarded Green's function we need the boundary action
\begin{eqnarray}
S &=&   \lim_{Z \ra \infty} - 2\wt{N} \int \frac{d\w dq}{(2\pi)^2} \D k_1 k_3 \inv{\nw^2} a_T'(Z,\nw) a_T(Z,-\nw) \nn \\
  &=&   \lim_{u \ra 0} - 4 \wt{N} (\pi T)^2 \int \frac{d\w dq}{(2\pi)^2}\inv{\nw^2} a_T'(u,\nw) a_T(u,-\nw) \nn \\
  &=&   - 4 \wt{N} (\pi T)^2 \int \frac{d\w dq}{(2\pi)^2}\inv{\nw^2} F'(0,\nw) F(0,-\nw) \nn \ ,
\end{eqnarray}
where we recovered $2\pi\a'$ and $\wt{N} \equiv N (2\pi\a')^2 = \frac{N_c N_f}{2(2\pi)^2}$.
The retarded Green function is then
\begin{eqnarray}
G^{\mathrm{ret}}_{xx} &=& \frac{\d^2 S}{\d a_x(0,\nw) \d a_x(0,-\nw)}  = \frac{\nw^2 \d^2 S}{\d F(0,\nw) \d F(0,-\nw)} \nn \\
  &=& - 8 \wt{N} (\pi T)^2 \frac{F'(0,\nw)}{F(0,\nw)} \ ,
\end{eqnarray}
and the conductivity is
\begin{eqnarray}
   \s_q&=& -\lim_{\w \ra 0} \frac{1}{\w} \mathrm{Im} G^{\mathrm{ret}}_{xx}(K) \Big|_{\w = |\vec{k}|} \nn \\
   &=& 8 \wt{N} (\pi T)^2 \lim_{\w \ra 0} \inv{\w} \mathrm{Im} \left( \frac{F_1'(0,\nw)\nw}{F_0} + \calo(\w^2) \right) \nn \\
   &=& \frac{N_cN_fT}{4\pi} \ .
\end{eqnarray}
Extending the procedure to finite density yields~\cite{MS}
\begin{eqnarray}
   \s_{q} = \frac{N_cN_fT}{4\pi}\sqrt{1+\nd^2} \ ,
\end{eqnarray}
where $\nd = \frac{{d}}{(\pi T)^3} = \frac{8n_q}{N_cN_f\sqrt{\l}T^3}$, which is (\ref{conductivity}) with $c=1$.

\section{D4/D8 Deconfined phase}

In this section we enforce an eigenmode analysis on the longitudinal
and transverse vector currents in the deconfined hot and dense D4/D8.
The eigenmode analysis parallels the one for cold and dense D4/D8 with
reflective boundary conditions at the BH horizon. No imaginary parts
arise from this analysis. In a way, in D4/D8 we may still entertain
the possibility of stationary solutions between the Left-pending and
Right-pending branes to mock up existing light bound states. Of course,
this is a formal suggestion.

The longitudinal operator ($\cald_L$) is
\begin{eqnarray}
  && \cald_L \equiv \dell_Z \frac{\sqrt{K(K-1)} \D^3}{\D^2 \w^2  - \frac{K-1}{K} q^2 } \dell_Z + \frac{K^{1/6}}{\sqrt{K-1}} \D \ .
\end{eqnarray}
When $q=0$ or $\w=0$ it is easily diagonalized, since
\begin{eqnarray}
    \cald_L(q=0) &=& \inv{\w^2} \dell_Z  \sqrt{K(K-1)} \D   \dell_Z + \frac{K^{1/6}}{\sqrt{K-1}} \D \ , \nn \\
    \cald_L(\w=0) &=& -\inv{q^2} \dell_Z  \frac{K^{3/2}}{\sqrt{K-1}} \D^3    \dell_Z + \frac{K^{1/6}}{\sqrt{K-1}} \D \ .
\end{eqnarray}
The Green's function ($\cald^{-1}_L$) may be expanded in terms of the complete set of eigenvalues
that diagonalize
\begin{eqnarray}
    && \cald_L(q=0) f  =  \left(\frac{K^{1/6}}{\w^2 \sqrt{K-1}} \D  \right) \l f\ ,   \\
    && \cald_L(\w=0) f  =  \left(\frac{K^{1/6}}{q^2 \sqrt{K-1}} \D  \right) \l f\ ,
\end{eqnarray}
where $  \frac{K^{1/6}}{\w^2 \sqrt{K-1}} \D  $ and  $ \frac{K^{1/6}}{q^2 \sqrt{K-1}} \D  $ are weight factors. Using
the complete sets,
\begin{eqnarray}
  && (\dell_Z \sqrt{K(K-1)} \D \dell_Z)\, \chi_n = - \left( \frac{K^{1/6}}{\w^2 \sqrt{K-1}} \D \right) \,  (\l^{\chi}_n)^2 \, \chi_n \ , \nn \\
  && \left(\dell_Z \frac{K^{3/2}}{\sqrt{K-1}} \D^3 \dell_Z\right)\, \xi_n = - \left(\frac{K^{1/6}}{\w^2 \sqrt{K-1}} \D \right) \,  (\l^{\xi}_n)^2 \, \xi_n \ , \nn
\end{eqnarray}
we have
\begin{eqnarray}
  && \bra{Z} \cald_{L}^{-1}(q=0) \ket{Z'} = \sum_{n \in \mathbb{N }} \frac{\chi_n(Z)
\chi_n(Z')}{-\w^2 + (\l^\chi_n)^2} + \frac{\chi_0(Z) \chi_0(Z')}{\w^2} \ ,  \\
  && \bra{Z} \cald_{L}^{-1}(\w=0) \ket{Z'} = \sum_{n \in \mathbb{N }} \frac{\xi_n(Z)
\xi_n(Z')}{q^2 + (\l^\xi_n)^2} + \frac{\xi_0(Z) \xi_0(Z')}{q^2} \ .
\end{eqnarray}

The transversal operator ($\cald_T$) is
\begin{eqnarray}
  && \cald_T \equiv -\inv{\w^2} \left(  \dell_Z \sqrt{K(K-1)} \D \dell_Z +  \frac{K^{1/6}}{\sqrt{K-1}}\left(\D \w^2 - \D^{-1}  \frac{K-1}{K} q^2 \right) \right) \ .
\end{eqnarray}
When $q=0$  it is easily diagonalized as
\begin{eqnarray}
    \cald_T (q = 0) \equiv -\inv{\w^2} \dell_Z \sqrt{K(K-1)} \D \dell_Z -\frac{K^{1/6}}{\sqrt{K-1}}  \D  \ .
\end{eqnarray}
With the eigenfunctions and eigenvalues:
\begin{eqnarray}
  && (\dell_Z \sqrt{K(K-1)}  \D \dell_Z)\, \zeta_n = - \frac{K^{1/6}}{\sqrt{K-1}} \D \,  (\l^{\zeta}_n)^2 \, \zeta_n \nn \ ,
\end{eqnarray}
the Green's function is expanded as
\begin{eqnarray}
   \bra{Z} \cald_{T}^{-1}(q=0) \ket{Z'} = \sum_{n \in \mathbb{N}} \frac{\zeta_n(Z)
\zeta_n(Z')}{\w^2 + (\l^\zeta_n)^2} + \frac{\zeta_0(Z) \zeta_0(Z')}{\w^2} \ .
\end{eqnarray}

\end{document}